\documentclass[Afour,sageh,times]{sagej_arxiv}

\usepackage{moreverb,url}

\usepackage[colorlinks,bookmarksopen,bookmarksnumbered,citecolor=red,urlcolor=red]{hyperref}

\newcommand\BibTeX{{\rmfamily B\kern-.05em \textsc{i\kern-.025em b}\kern-.08em
T\kern-.1667em\lower.7ex\hbox{E}\kern-.125emX}}

\newcommand{\dx}{\dftl{x}}

\newcommand{\dt}{\dftl{t}}

\newcommand{\be}{\begin{equation}}
\newcommand{\ee}{\end{equation}}
\newcommand{\bes}{\begin{equation*}}
\newcommand{\ees}{\end{equation*}}
\newcommand{\bse}{\begin{subequations}}
\newcommand{\ese}{\end{subequations}}

\def\ee{{\hat {\underline e}}}

\def\dt{ \Delta t }
\def\dx{ \Delta x }

\def\scriptO{{{\it O}\kern -.42em {\it `}\kern + .20em}}
\def\RR{{{\rm l}\kern - .15em {\rm R} }}
\def\PP{{{\rm l}\kern - .15em {\rm P} }}
\def\L2{{{\sf L}^2}}
\def\H1{{{\sf H}^1}}

\def\PN2{{\PP_{N}-\PP_{N-2}}}

\def\complex{{{\rm C} \kern - .53em {\rm l} \kern + .38em}}

\def\a1{{ | \lambda_{\min} |}}

\def\l1{{   \lambda_{\min}  }}

\def\bu0{{\underline {\bf 0}}}

\def\bu{{\bf u}}

\def\u0{{\underline 0}}
\def\n12{{n_{\frac{1}{2}}}}
\def\t12{{t_{\frac 1 2}}}
\def\n12{{n_{0.8}}}
\def\t12{{t_{0.8}}}

\newcommand{\pp}[2]{\frac{\partial #1}{\partial #2} }

\begin{document}

 \runninghead{}
 \title{Towards Exascale for Wind Energy Simulations} 

 \author{Misun Min\affilnum{1}, 
        Michael Brazell\affilnum{2},
        Ananias Tomboulides\affilnum{3},
        Matthew Churchfield\affilnum{4},
        Paul Fischer\affilnum{1,5,6} and
        Michael Sprague\affilnum{4}}

 \affiliation{\affilnum{1}Mathematics and Computer Science, 
                         Argonne National Laboratory, Lemont, USA\\
             \affilnum{2}Computational Science Center,
                         National Renewable Energy Laboratory, Golden, USA\\
             \affilnum{3}Mechanical Engineering, 
                         Aristotle University of Thessaloniki, Tessaloniki, Greece\\
             \affilnum{4}National Wind Technology Center,
                         National Renewable Energy Laboratory, Golden, USA\\
             \affilnum{5}Computer Science,
                         University of Illinois Urbana-Champaign, Urbana, USA\\
             \affilnum{6} Mechanical Science \& Engineering,
                         University of Illinois Urbana-Champaign, Urbana, USA}

 \corrauth{Misun Min, 
           Mathematics and Computer Science, 
           Argonne National Laboratory, Lemont, USA}
 \email{mmin@mcs.anl.gov}

 \begin{abstract}
  We examine large-eddy-simulation modeling approaches and computational
performance of two open-source computational fluid dynamics codes for the
simulation of atmospheric boundary layer (ABL) flows that are of direct
relevance to wind energy production.
The first is NekRS, a high-order, unstructured-grid, spectral element code.
The second, AMR-Wind, is a block-structured, second-order finite-volume 
code with adaptive-mesh-refinement capabilities.
The objective of this study is to co-develop these codes in order to improve
model fidelity and performance for each. These features will be critical for
running ABL-based applications such as wind farm analysis on advanced computing
architectures.  To this end, we investigate the performance of NekRS and AMR-Wind
on the Oak Ridge Leadership Facility supercomputers Summit, using 4 to 800 nodes
(24 to 4,800 NVIDIA V100 GPUs), and Crusher, the testbed for the Frontier exascale system using 18 to 384 Graphics Compute Dies on AMD MI250X GPUs.  
We compare strong- and weak-scaling capabilities,
linear solver performance, and time to solution.  We also identify leading
inhibitors to parallel scaling.

 \end{abstract}


 \maketitle


 \section{Introduction}
\label{sec:intro}

Atmospheric boundary layer (ABL) flows are an important part of everyday life.  
Aside from being a primary driver of vertical exchanges in moisture, aerosols, 
and atmospheric gases, the ABL affects practical aspects of life---including   
the transportation system, 
renewable energy generation, pollution dispersion, noise propagation, and 
transmission of electromagnetic signals.
ABL flows are turbulent, and the state of the turbulence is affected by density 
stratification that arises in large part from surface heating and cooling.  
Additionally, Coriolis effects caused by planetary rotation and curvature 
complicate the flow.  Furthermore, regional-scale weather patterns and terrain 
add complexity to the ABL.
Significant research effort is applied to ABL flows because of their importance 
and complexity~\cite{Moeng:1984,
Berg-Kelley:2020,
beare2006,
Sullivan-etal:2008,
kosovic2000,
Pedersen:2014,
Mirocha:2020,
Churchfield-Moriarty:2020}.
This work focuses on numerical computation of ABL flows using 
large eddy simulation (LES), where the governing physics equations are solved in 
filtered form such that the larger, energy-containing eddies are directly resolved, 
and the remaining ``subgrid-scale'' (SGS) turbulence is modeled. LES was born out 
of ABL research roughly five decades ago~\cite{Lilly:1962,Smagorinsky:1963}, 
and continues to evolve and improve.

Wind energy is a prime example of an application driven by the ABL.  Generation
of electrical energy from farms of wind turbines at night in the stable ABL is
a particularly interesting situation.  The winds tend to be stronger, so
generation is higher.  With decreased turbulence, wind turbine wakes persist
for longer distances, significantly affecting wind farm efficiency and fatigue
loads on waked wind turbines.  With this example in mind, researchers wish to
increase grid resolution to reduce reliance on the SGS turbulence model, but
they also wish to increase the overall domain size to encompass the
wind farm, which commonly extends many kilometers horizontally.  Increased
domain size is desirable in many other applications besides wind energy.  For
example, LES can be used to study deep convection, which happens over a large
geographical extent many kilometers into the atmosphere, and there is a push
toward LES of regional-scale weather.

High-fidelity LES of the turbulent ABL is dependent on massively parallel
high-performance computing (HPC).  HPC architectures are evolving from
traditional homogeneous x86-CPU-based computing.  For example, the world's
second-fastest computer (as of June 2022), the Supercomputer Fugaku at the RIKEN
Center for Computational Science~\cite{fugaku}, is built around Fujitsu's
custom ARM A64FX processor and does not use a GPU.  Alternatively, the U.S.
Department of Energy (DOE) has embraced a hybrid CPU-GPU approach for
its leadership-class computing. 
Summit~\cite{summit}, the world's fourth-fastest computer, has nodes that house
two IBM POWER9 CPUs (each with 22 cores) and six Nvidia V100 GPUs and is
capable of $200 \times 10^{15}$ floating-point operations per second (FLOPS).
Similarly, DOE's first exascale-class supercomputers, i.e., those capable
of at least $10^{18}$ FLOPS, will be a hybrid CPU-GPU based systems. 
Frontier, the world's first exascale class supercomputer at the 
Oak Ridge Leadership Computing facility, has nodes that house one 
AMD CPU and 4 AMD MI250X GPUs~\cite{frontier}.

\begin{figure*}[!t]
  \begin{center}
   \includegraphics[width=0.9\textwidth]{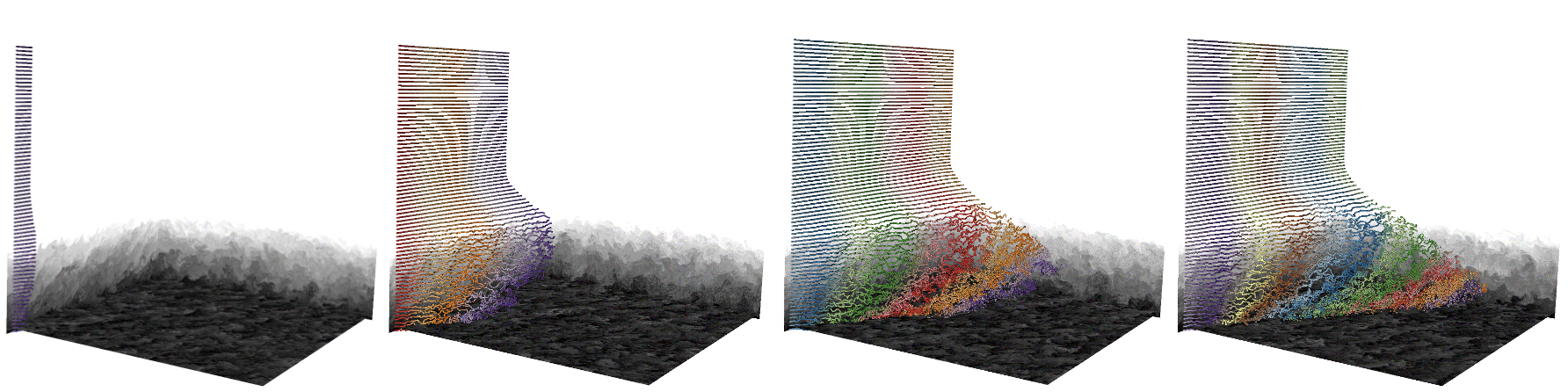}
  \end{center}
  \caption{\label{abl-demo}
  Illustration for an atmospheric boundary layer simulation with tracer particles
    for the GABLS benchmark problem. (Simulation by N. Lindquist \cite{lindquist21}.)
}
\end{figure*}

As described in a 2015 DOE workshop report~\cite{Sprague-etal:2017}, the
transition to exascale computing brings many opportunities in computational
fluid dynamics (CFD), as well as significant challenges.  Two of the grand
challenge opportunities described in that workshop report are relevant to this
paper:  the simulation of {\em boundary layer turbulence over large areas} and
the simulation of {\em an entire wind farm under realistic atmospheric flow
conditions}.  The transition to GPU-accelerated computing is significant for
those performing  CFD for numerical weather forecasting or for LES of ABL
flows.  While hybrid CPU-GPU processing potentially enables dramatically faster
computing (at low power), legacy CPU-based codes need significant overhauls or
rewrites to run effectively in a CPU-GPU
environment~\cite{Sprague-etal:2017,Robinson-Sprague:2020}.

Several groups have introduced CFD codes for LES of atmospheric flows with an
emphasis on GPU calculations.  While early efforts in weather forecasting 
on GPUs focused at $O(100)$ m resolution (see, e.g.,~\cite{Schalkwijk:2015}),
more recent efforts have performed high-fidelity GPU-based LES on $O(1)$ m grid
sizes. 
Van Heerwaarden et al.~\cite{VanHeerwaarden-etal:2017} introduced
the MicroHH 1.0 incompressible-flow solver directed at atmospheric flow; the
solver is based on finite-difference spatial discretization and a
split-operator time integration.  The authors showed that for problem sizes
that fit, a single GPU had performance similar to that of 32 CPU cores.
Sauer and M\~{u}noz-Esparza~\cite{Sauer-MunozEsparza:2020} introduced
the FastEddy LES model that was created for CPU and GPU systems. FastEddy
solves the fully compressible Navier--Stokes equations using finite-difference
spatial discretization and explicit Runge--Kutta time integration. The authors
showed excellent scaling on up to 32 GPUs and argued that one GPU provides
performance similar to that of 256 CPU cores.  
Recent high-order incompressible flow codes using fast tensor-product operator
evaluation include ExaDG \cite{arndt2020} and SPECHT\_FS \cite{stiller_ns_2019} and
deal.ii \cite{krank2017}.  ExaDG and deal.ii use a discontinuous Galerkin
formulation, whereas SPECHT\_FS uses a continuous Galerkin formulation similar to
that in Nek5000/RS.

In this paper we examine LES modeling approaches and computational performance
of two open-source, incompressible-flow, GPU-oriented CFD codes that employ
fundamentally different spatial discretization and data structures.  The first
is NekRS~\cite{nekrs0,nekrs,nekrs1}, which is an updated version of the
Nek5000 code~\cite{nek5000}.  Nek5000/RS is a high-order, unstructured-grid,
spectral-finite-element CFD code.  The second is AMR-Wind~\cite{amrwind}, which
is part of the ExaWind simulation suite~\cite{Sprague-etal:2020}.  AMR-Wind is
a block-structured, second-order, finite-volume-method CFD code with adaptive
mesh refinement (AMR) capabilities and is built on the AMReX library, a
software framework for massively parallel, block-structured
applications~\cite{amrex}.  Both of these codes are part of the U.S. DOE Exascale
Computing Project, which is supporting the development of GPU-ready
applications for exascale-class
supercomputers~\cite{ecp}\cite{Alexander-etal:2020}.

We compare NekRS and AMR-Wind predictions and performance on the
well-known GEWEX (Global Energy and Water Cycle Experiment) Atmospheric
Boundary Layer Study (GABLS) stably stratified benchmark LES
case~\cite{beare2006}, which is illustrated in Fig.~\ref{abl-demo}.  The flow
comprises a quiescent mean flow above $\approx 200$~m (going to the right and
into the page in Fig.~\ref{abl-demo}) with a sheared turbulent boundary layer 
(going to the right and out of the page) over the 0--200~m vertical range.
The flow is coupled with thermal buoyancy effects.
%
The computational domain is 400~m $\times$ 400~m $\times$ 400~m and doubly periodic in
the streamwise ($x$) and cross-flow ($y$) directions.  Potential temperature
distributions at Reynolds number $Re=50M$ and $t$=6 hours are illustrated in
Fig.~\ref{fig:512} along with profiles in Fig. \ref{fig:mean}.

In addition to investigating these codes' ability to represent ABL physics, an
objective of this study is to co-develop these codes in order to improve model
fidelity and performance, features that will be critical for running ABL-based
applications such as wind farm analysis on advanced computing architectures.
By careful cross-comparison, both codes have made significant advances.  This
article focuses on performance.  A separate article addresses subgrid-scale
modeling for LES of the ABL.
Here we investigate the scalability of NekRS and AMR-Wind on the Oak Ridge
Leadership Computing Facility supercomputer Summit, using 4 to 800 nodes 
(24 to 4,800 NVIDIA V100 GPUs). We provide iteration counts, average-time 
per step, and the real-time ratio (ratio of wall-clock time to physical time),
as well as detailed performance metrics.   We additionally include a limited
set of timing data for the two codes on Crusher, using up to
384 Graphics Compute Dies (GCDs) on AMD MI250X GPUs (one MPI rank per GCD).

The paper is organized as follows.
Section II describes the codes and gives an overview of the numerical approaches used.
Section III discusses the numerical setup of the  simulations.
Section IV provides studies comparing the codes'  performance and scaling.
Section V presents a brief summary.

 \section{Description of Codes}
\label{sec:codes}

The numerical results are based on LES, which requires enhanced dissipation 
to provide an energy drain at the grid scale.
Consequently, the incompressible Navier--Stokes (NS) and  potential temperature
equations are solved in a {\em spatially filtered} resolved-scale formulation,
expressed in nondimensional form as
\index{Navier-Stokes}
\begin{eqnarray}
    \label{eq:ns}
    \pp{\bar u_i}{t} + \bar u_j\frac{\partial {\bar u_i}}{\partial x_j}
     &=& -\frac{1}{\bar \rho}\frac{\partial {\bar p}}{\partial x_i}
         -\pp{\tau_{ij}}{x_j} + f_i - \frac{\theta^{\prime}}{\theta_0} g_i, \\
    \label{eq:inc}
    \frac{\partial {\bar u_j}}{\partial x_j} &=& 0, \\
    \label{eq:energy}
    \pp{\bar \theta}{t} + {\bar u_j}\frac{\partial {\bar \theta}}{\partial x_j}
     &=& -\pp{\tau_{\theta j}}{x_j}\,,
\end{eqnarray}
where an overbar denotes the LES filtering operation such that
$\bar u_i$ is the $i$th component of the resolved-scale velocity vector, $\bar \rho$ is
the density,  $\bar p$ is the pressure, $g_i$ is the gravity acceleration vector,
and $\bar \theta$ is the potential temperature in the resolved scale. 
The scalar $\theta^{\prime}/\theta_0$ that dictates the sign and strength of the
buoyancy force is obtained from
\begin{eqnarray}
 \frac{\theta^{\prime}}{\theta_0} = \frac{\bar{\theta}-\theta_0}{\theta_0},
\end{eqnarray}
where $\theta_0$ is the reference potential temperature and $f_i$ includes the Coriolis acceleration.
In addition, $\tau_{ij}$ and $\tau_{\theta j}$ are the stress tensors in the momentum
and energy equations, respectively, which include (and are dominated by) SGS modeling terms
\begin{equation}
 \tau_{ij} = -\frac{2}{Re} S_{i j} + \tau^{sgs}_{ij} = 
   -\frac{1}{Re} \left( \frac{\partial{\bar u_i}}{\partial x_j} + 
   \frac{\partial {\bar u_j}}{\partial x_i} \right) + 
   \tau^{sgs}_{ij}, 
\label{eqn.tauij}
\end{equation}
\noindent
and
\begin{equation}
   \tau_{\theta j} = -\frac{1}{Pe} \frac{\partial{\bar \theta}}{\partial x_j} + 
   \tau^{sgs}_{\theta j}, 
\end{equation}
where $Re$ is the Reynolds number, $Pe$ is the Peclet number, 
$S_{ij}$ is the resolved-scale strain-rate tensor and
$\tau^{sgs}_{ij}$ and $\tau^{sgs}_{\theta j}$ are the subgrid-scale stress tensors.

The SGS modeling in Nek5000/RS is based on~\cite{sullivan94}, where the SGS
stress tensors $\tau^{sgs}_{ij}$ and $\tau^{sgs}_{\theta j}$ are expressed in
terms of a non-isotropic, mean-field eddy viscosity (MFEV) obtained by the
horizontally averaged mean strain rate, and an isotropic, fluctuating part.
The SGS model of~\cite{sullivan94} is based on the following expression:
\begin{equation}
\tau^{sgs}_{i j}=-2 \nu_{t} \gamma S_{i j}-2 \nu_{T}\left\langle S_{i j}\right\rangle,
\label{eqn.tijdef}
\end{equation}
\noindent
where the angle brackets $\left\langle \ \ \right\rangle$ denote averaging over the
homogeneous directions and $\nu_T$ is an average eddy viscosity that is expressed in 
terms of mean flow quantities. In Eq.~(\ref{eqn.tijdef}) $\gamma$ is
an ``isotropy factor,'' which accounts for variability in the SGS constants due to 
anisotropy of the mean flow. In~\cite{sullivan94}, the fluctuating eddy viscosity, 
$\nu_t$, is obtained by using an eddy viscosity model based on the SGS turbulent kinetic 
energy equation, in which the shear production term is computed from the fluctuating 
velocities as suggested by~\cite{Schumann1975}.

Here, the fluctuating (isotropic) part is taken into account through the use of either a 
high-pass filter (HPF)~\cite{Stolz_Schlatter_2005} or a Smagorinsky (SMG) model based on the 
fluctuating strain rate. For the former model, which is not eddy-viscosity based, $\nu_t$ in 
Eq.~(\ref{eqn.tijdef}) is by definition equal to zero.
On the other hand, the expression for $\nu_T$ is derived so that the law-of-the-wall behavior
can be recovered in the absence of any resolved turbulence.
 
The SGS modeling in AMR-Wind is based on~\cite{Smagorinsky:1963}, where a single
partial differential equation for subgrid-scale kinetic energy is solved, and
from that a subgrid-scale eddy viscosity is computed.  The Boussinesq eddy
viscosity hypothesis is then invoked to obtain the subgrid-scale stress tensor
and heat flux vector.

In the following subsections we discuss the details of the numerical approaches
of Nek5000/RS and AMR-Wind. For simplicity, we use $u_i$, $p$, and $\theta$,
dropping the overbar notation from $\bar u_i$, $\bar p$, and $\bar \theta$ 
in the remaining sections.

\subsection{Nek5000/RS}

Nek5000~\cite{nek5000} is a spectral element code that is used for a wide range
of thermal-fluids applications.
It employs high-order spectral elements
\cite{pat84} in which the solution,
data, and test functions are represented as {\em locally structured} $N$th-order
tensor-product polynomials on a set of $E$ {\em globally unstructured}
curvilinear hexahedral brick elements. The approach yields two principal
benefits.  First, for smooth functions such as solutions to the incompressible
NS equations, high-order polynomial expansions exhibit rapid convergence with
approximation order, often yielding a significant reduction in the number of
unknowns ($n \approx EN^3$) required to reach engineering tolerances.  Second,
the locally structured forms permit local lexicographical ordering with minimal
indirect addressing and, crucially, the use of tensor-product sum factorization
to yield low $O(n)$ storage costs and $O(nN)$ work complexities \cite{sao80}.

NekRS \cite{nekrs} is a GPU-accelerated version of Nek5000 that is targeting
high performance on forthcoming exascale platforms.  For performance
portability, NekRS is written in C++/OCCA~\cite{occa}.  Several key kernels
are based on highly tuned OCCA kernels coming from the development work
of Warburton and co-workers in the libParanumal library~\cite{libp}.  Specific
attention in NekRS has been given to ensure scalability to $P=10^4$--$10^5$
ranks and beyond \cite{gb21}.  NekRS retains access to the standard Nek5000
interface, which allows users to leverage existing user-specific source code
such as statistical analysis tools for turbulence.

Time integration in Nek5000/RS is based on a semi-implicit splitting scheme
using $k$th-order backward differences (BDF$k$) to approximate the time
derivative coupled with implicit treatment of the viscous and pressure
terms and $k$th-order extrapolation (EXT$k$) for the remaining advection and forcing
terms.  This approach leads to independent elliptic subproblems comprising a
Poisson equation for the pressure, a coupled system of Helmholtz equations for
the three velocity components, and an additional Helmholtz equation for the
potential temperature. 
The pressure Poisson equation is obtained by taking the
divergence of the momentum equation and forcing $\frac{\partial u_i^n}{\partial
x_i}=0$ at time $t^n=n \Delta t$. The velocity and temperature
Helmholtz equations are obtained once $p^n$ is known:
\begin{eqnarray} \label{pres}
-\frac{\partial^2}{\partial x_j\partial x_j} p^n\! &=&\! q^n, 
\\ \label{vel}
\frac{\beta_0}{\dt} u_i^n 
-\pp{}{x_j}\left(\frac{1}{Re}+\gamma \nu_t \right) 2S^n_{ij}
                              \! &=&\! - \pp{}{x_i} p^n + r_i^n, 
\\ \label{temp}
\hspace*{-.1in}
\left[\frac{\beta_0}{\dt} -\pp{}{x_j}\left(\frac{1}{Pe}+ 
   \gamma \nu_t\right) \pp{}{x_j}\right] \theta^n \! &=&\!  s^n,
\end{eqnarray}
where $\beta_0$ is an order-unity constant associated with BDF$k$
\cite{fischer17}; $S_{ij}$ is the resolved-scale strain-rate tensor as
described in~(\ref{eqn.tauij}); and $q^n$, $r_i^n$, and $s^n$ represent the sum
of the values from the previous timesteps for the contributions from BDF$k$ and
EXT$k$.  Also included in $r_i^n$ and $s^n$ are eddy diffusivity terms coming
from the mean-field eddy diffusivity mentioned above.  The fully coupled system
of Helmholtz equations for the three velocity components~(\ref{vel}) is used
only when the fluctuating (isotropic) part of the SGS stress tensor is modeled
using an SMG model based on the fluctuating strain rate. When this
part is modeled through the use of a high-pass filter
(HPF)~\cite{Stolz_Schlatter_2005}, the resulting Helmholtz equations for the
three velocity components are not coupled. 


With the given time-splitting, we recast (\ref{pres})--(\ref{temp}) into weak
form and derive the spatial discretization by restricting the trial and test
spaces to be in the finite-dimensional space spanned by the spectral element
basis.   The discretization leads to a sequence of symmetric positive definite
linear systems for pressure, velocity, and temperature.  Velocity and temperature 
are diagonally dominant and readily addressed with Jacobi-precondition conjugate
gradient iteration.  The pressure Poisson solve is treated with GMRES using
$p$-multigrid as a preconditioner.  Details of the formulation can be found in
\cite{fischer04,fischer17,malachi2022a}.


NekRS supports several features to accelerate performance, including overlapped
communication and computation during operator evaluation, which yields a
10--15\% performance gain for NS simulations; FP32 local-operator inversion and
residual evaluation for the Chebyshev-accelerated Schwarz-based $p$-multigrid;
and projection of the velocity and pressure solutions onto the space
of prior solutions to generate an initial guess, which can yield
a 1.5--2-fold NS performance gain \cite{fisc98}.
On the NVIDIA A100, the OCCA-based kernels are close to the bandwidth-limited
roofline and are sustaining 2.1--2.2 TFLOPS (FP64) for the Poisson operator,
and 3.1--3.8 TFLOPS (FP64) for the advection operator.
  In the pressure preconditioner, the forward Poisson operator on the coarser
multigrid levels realizes 2.5--3.9 TFLOPS (FP32) and the Schwarz smoother
sustains 2.5--5.1 TFLOPS (FP32).  (The lower values are for smaller values of
$N$ that are used in the $p$-multigrid V-cycle.)   Comparable values are realized
on the NVIDIA V100s on Summit, save that they are $\approx$ 1.5 times lower
than those on the A100.  
Sustained flop rates for the full NS solver are $\approx 470$ GFLOPS
per V100 on Summit, as discussed in Section \ref{sec:tune}.



\subsection{AMR-Wind}

AMR-Wind~\cite{amrwind} is a spatially and temporally second-order accurate 
finite-volume code. Important aspects of the discretization are discussed below; 
for more details readers can see \cite{Almgren1998} since the discretization 
is similar to the incompressible-flow solver IAMR~\cite{iamr}. 
Velocity, scalar quantities, and gradients of pressure are located
at cell centers, whereas pressure is located at nodes. Partial staggering
combined with an approximate projection method yields linear systems that are
well studied, have small-bandwidth stencils, and can be efficiently solved with
standard techniques such as geometric multigrid. These discretization choices
give a well-balanced mix of both efficiency and accuracy. In addition to the
spatial staggering there is also staggering in time similar to a Crank--Nicolson
formulation. The time discretization is
\begin{align}
 \frac{c_k^{n+1}- c_k^n}{\Delta t} &+ \left[ \frac{\partial {c}_k {u}_j }{\partial x_j}  \right]^{n+1/2} \nonumber\\
 &=\frac{1}{\rho^{n+1/2}} \frac{\partial q^{n+1}_{kj}}{\partial x_j} + G_k^{n+1/2}, \\
  \frac{u_i^* - u_i^n}{\Delta t} &+ \left[ \frac{\partial {u}_i {u}_j }{\partial x_j}  \right]^{n+1/2} \nonumber\\
 &= \frac{1}{\rho^{n+1/2}} \left( \frac{\partial \tau^{n+1}_{ij}}{\partial x_j} -\frac{\partial p}{\partial x_i}^{n-1/2}\right)  + F_i^{n+1/2},
\label{eqn:ns-disc}
\end{align}
where $c$ indicates a scalar quantity and $u$ denotes velocity.  
The index $n$ represents a time step, the index $i$ runs over the three
momentum equations, the index $k$ runs over the scalar equations, and the
repeated index $j$ indicates summation. $F_i^{n+1/2}$ and $G_k^{n+1/2}$ are source terms and are evaluated at time step $n+1/2$.
 $\rho$ is density, which in these simulations is constant in time and 
space, but we leave the time level to show at what point in time density is 
evaluated if the code is run in variable-density mode.

The advection term is formed by extrapolating in time by using a Godunov method
\cite{Almgren1998}. Specifically, the velocity is first extrapolated in space (to
the faces) and in time to $n+1/2$ in a predictor step. 
MAC projection~\cite{Bell1991AnES} is applied to ensure that the face velocities are divergence free, which takes the form
\[ P^{\mathrm{MAC}}(u_i^f) =
u^f_i -  \frac{1}{\rho^{n}} \left(  \frac{\partial \psi}{\partial x_i} \right),
\]
where $u_i^f$ represents a face velocity and $\psi$ is a Lagrange multiplier
located on the cells. Setting the divergence equal to zero forms
a variable coefficient Poisson equation:
\[ \frac{\partial}{\partial x_j} \left( \frac{1}{\rho^{n}} \frac{\partial \psi}{\partial x_j} \right) = \frac{\partial u_j^f}{\partial x_j}. \]
The Poisson equation is discretized by using a cell-centered seven-point stencil
in 3D, which is efficiently solved by using multilevel multigrid as a linear solver
\cite{Weiqun2019}.
Once the velocity on the faces is divergence free, the advection terms are
formed. Options for discretizing these advection terms include Godunov PLM
\cite{Vanleer:1977}, PPM \cite{Colella:1984}, and WENO-Z \cite{motheau2020}.
The Godunov schemes are high-order accurate and use an extended stencil.
The scalar equations (e.g., potential temperature) are advanced one at a
time by solving a Helmholtz problem. This Helmholtz problem is discretized
by using a cell-centered finite-difference method forming a 7-point stencil in 3D
\cite{Almgren1998}. The momentum equations are saved for last to allow the
source terms to be evaluated at $n+1/2$ by using the previously updated scalar
equations. A good example is the Boussinesq buoyancy term, which adds a
source term at time step $n+1/2$ to the momentum equation. The Boussinesq
buoyancy term is a function of the already advanced potential temperature, which
is averaged to time $t^{n+1/2}=(n+1/2)\Delta t$. The scalar equations and momentum equations can
be solved by using geometric multigrid or, if diagonally dominant enough, a
Krylov method such as {\tt bicgstab} (biconjugate gradient stabilized) is sufficient. 
In this work, because of the small
time step, we use only {\tt bicgstab} in all of the Helmholtz solves.

The intermediate velocity $u^*_i$ is advanced by solving a Helmholtz problem in
tensor form. However, this velocity vector  $u_i^*$ is not guaranteed to be
divergence free. An approximate projection method is used to solve for the
velocity at time $t^{n+1}=(n+1)\Delta t$:
\begin{equation}
u^{n+1} = P(u_i^*)\,,
\label{eq:projection1}
\end{equation}
where the nodal projection $P$ is defined to be
\begin{equation}
 P(u_i^*) = u_i^* + \frac{\Delta t}{\rho^{n+1/2}} \left( \frac{\partial p}{\partial x_i}^{n-1/2} 
 -  \frac{\partial \phi}{\partial x_i} \right)\,.
 \label{eq:projection2}
\end{equation}
This approximate projection is different from the algorithm in
\cite{Almgren1998} and more similar to the projection in \cite{Almgren:2000}
labeled version 2.  Taking the divergence of (\ref{eq:projection2}) and
setting it equal to zero, we have
\begin{equation}
\frac{\partial}{\partial x_j} \left( \frac{\Delta t}{\rho^{n+1/2}} \frac{\partial \phi}{\partial x_j} \right) = \frac{\partial}{\partial x_j} \left( u_j^* + \frac{\Delta t}{\rho^{n+1/2}} \frac{\partial p}{\partial x_j}^{n-1/2} \right)\,,
\label{eq:projection3}
\end{equation}
where $\phi$ is a Lagrange multiplier related to the pressure field and solved on
the nodes. To solve the nodal projection in~(\ref{eq:projection3}), a
variational form is used, which leads to a 27-point stencil \cite{Almgren1998}.
The right-hand side of~(\ref{eq:projection3}) is formed by taking finite
differences across cells and averaging them to each node. Each node has four
contributions in each coordinate direction. Once the solution $\phi$ is
obtained, the velocity is updated by using~(\ref{eq:projection2}), and the
pressure and its gradient are updated by using
\[ p^{n+1/2} \leftarrow \phi \quad,
\frac{\partial p}{\partial x_i}^{n+1/2} \leftarrow \frac{\partial
\phi}{\partial x_i}\,.  \]
The gradient $\partial\phi /\partial x_i$ is approximated by taking finite
differences of $\phi$ along edges and averaging each edge to the cell center.
For each coordinate direction four edges contribute to the value
at the cell center.

\begin{figure*}
 \footnotesize
  \begin{center}
     \includegraphics[width=0.22\textwidth]{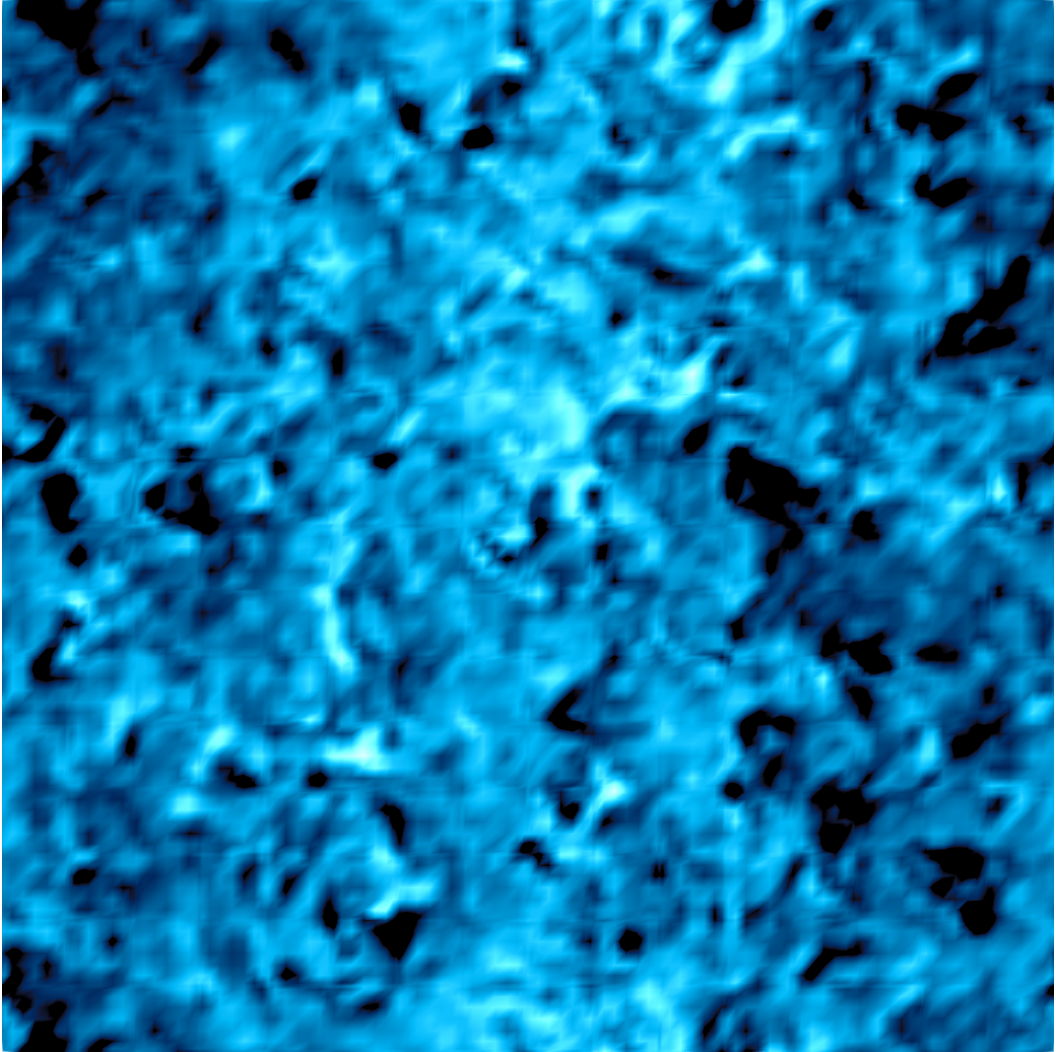}
     \includegraphics[width=0.22\textwidth]{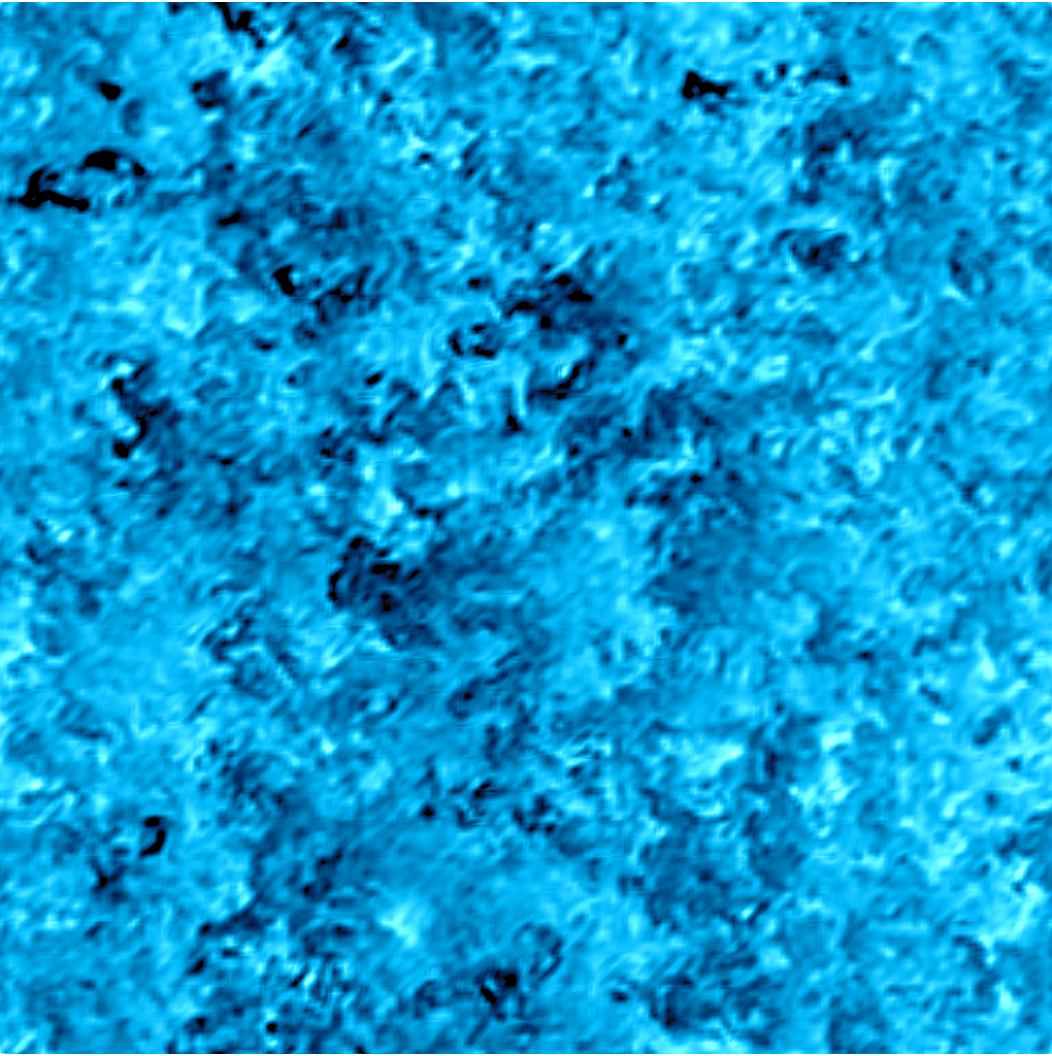}
     \includegraphics[width=0.22\textwidth]{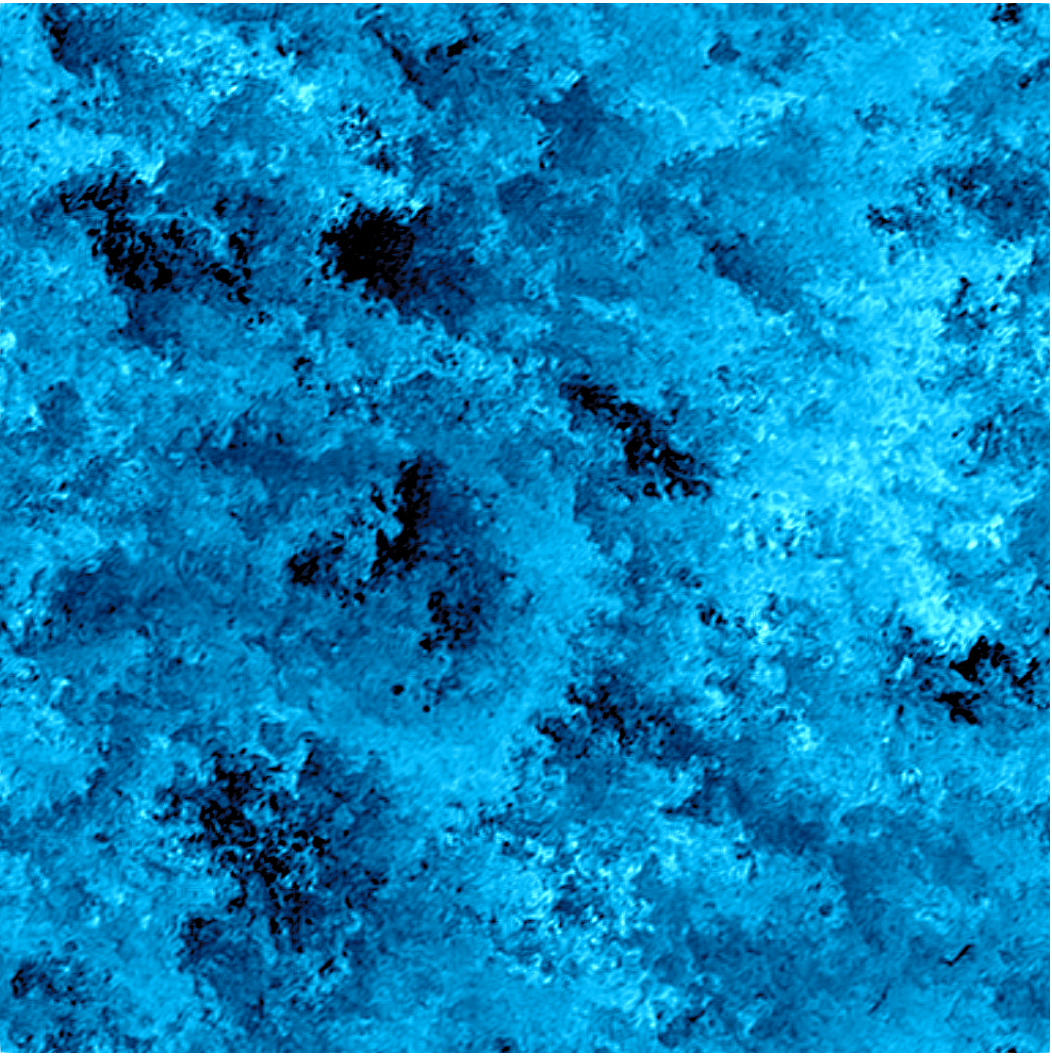}
     \includegraphics[width=0.04\textwidth]{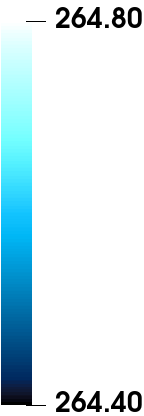}
     \hspace{-4em}\\
     \footnotesize{NekRS(HPF), $z$= 100 m, $\Delta x$= 3.12 m, 1.56 m, 0.78 m ($n=128^3, 256^3, 512^3$)} \\
     \includegraphics[width=0.22\textwidth]{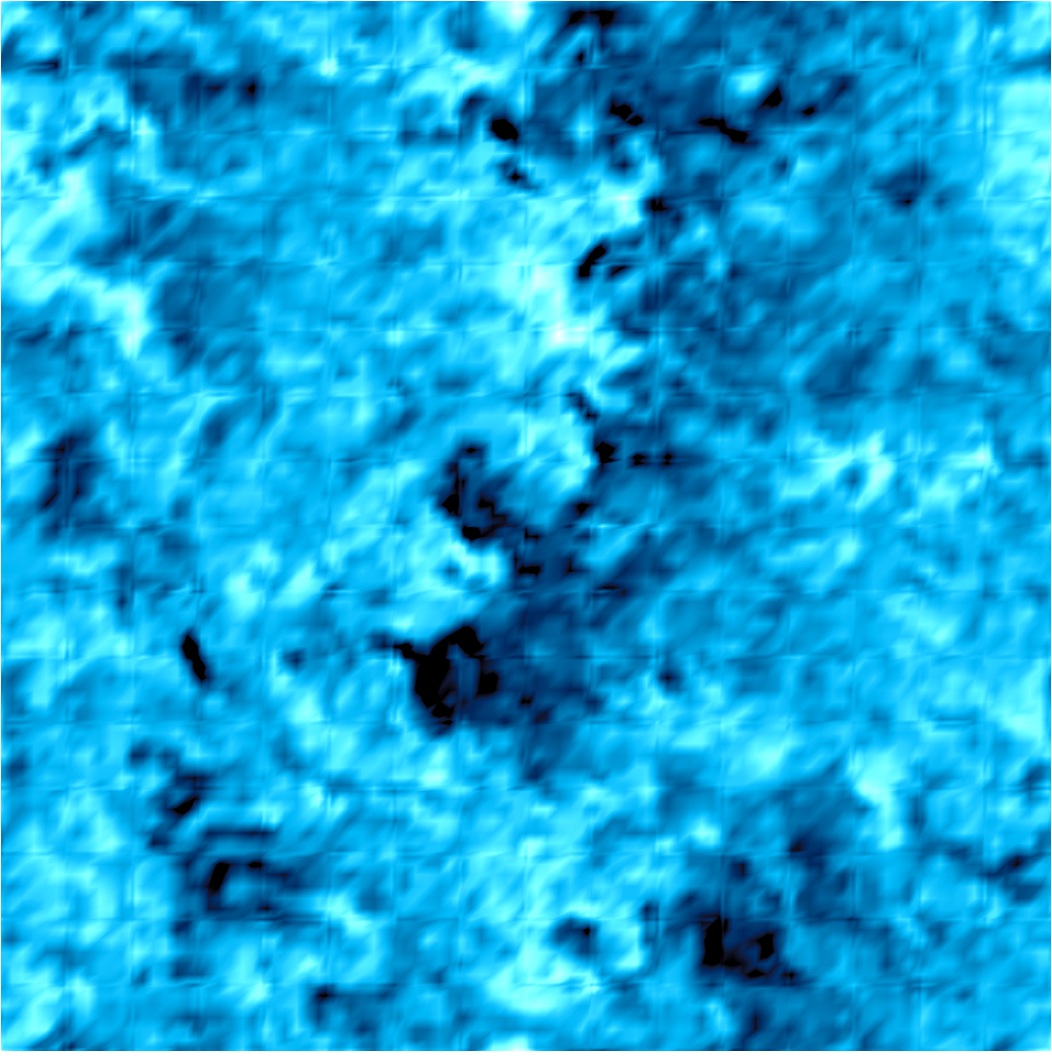}
     \includegraphics[width=0.22\textwidth]{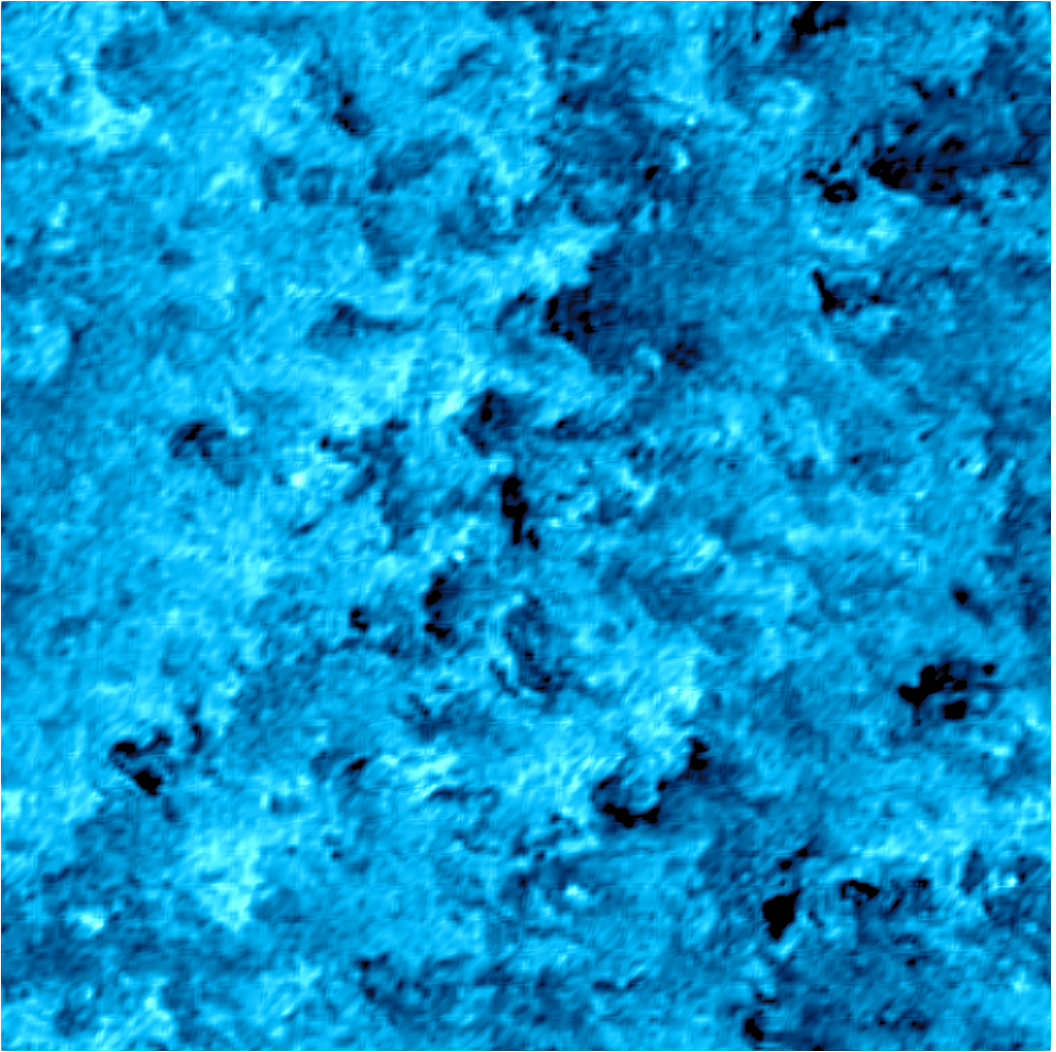}
     \includegraphics[width=0.22\textwidth]{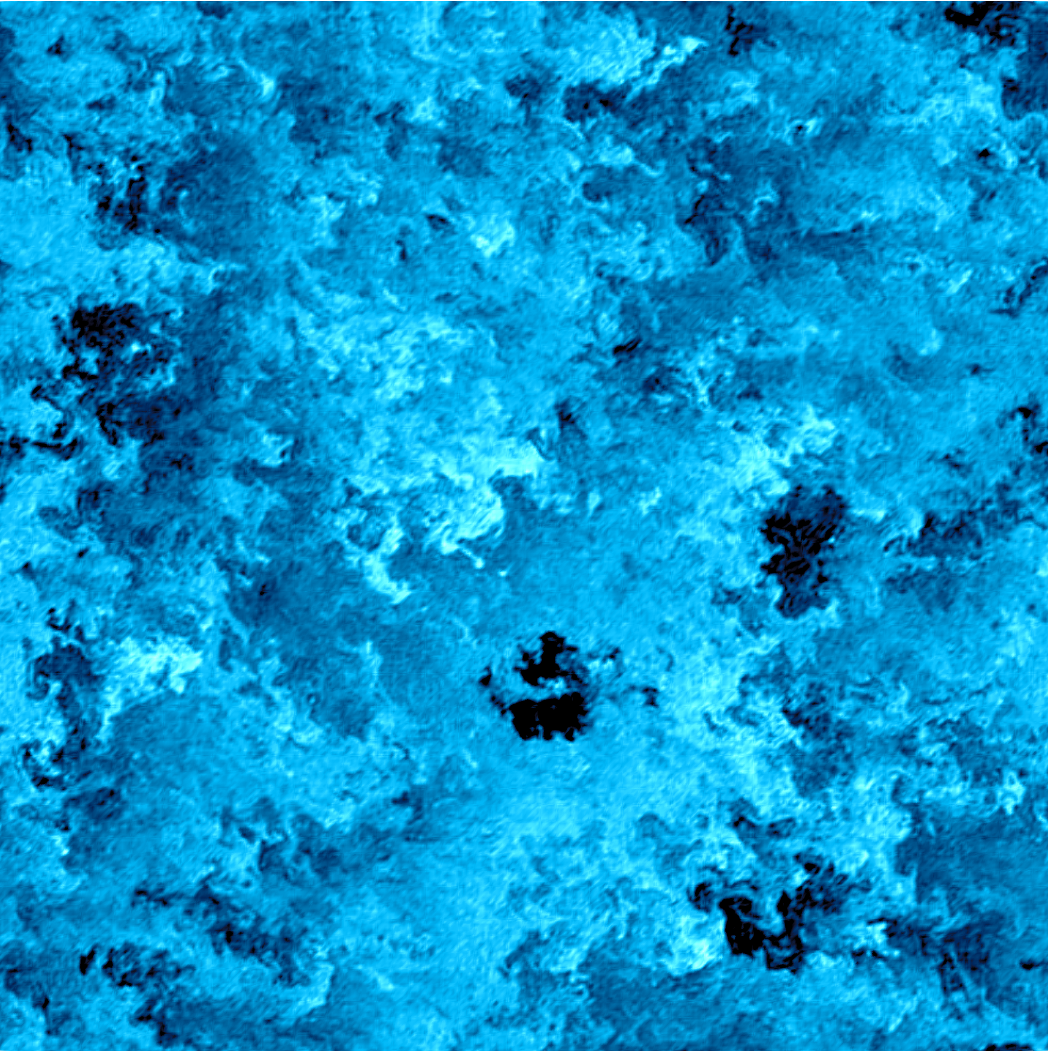}
     \includegraphics[width=0.04\textwidth]{./figs/bar-n512-z100.png}
     \hspace{-4em}\\
     \footnotesize{NekRS(SMG), $z$= 100 m, $\Delta x$= 3.12 m, 1.56 m, 0.78 m ($n=128^3, 256^3, 512^3$)} \\
     \vskip.1in
     \includegraphics[width=0.22\textwidth]{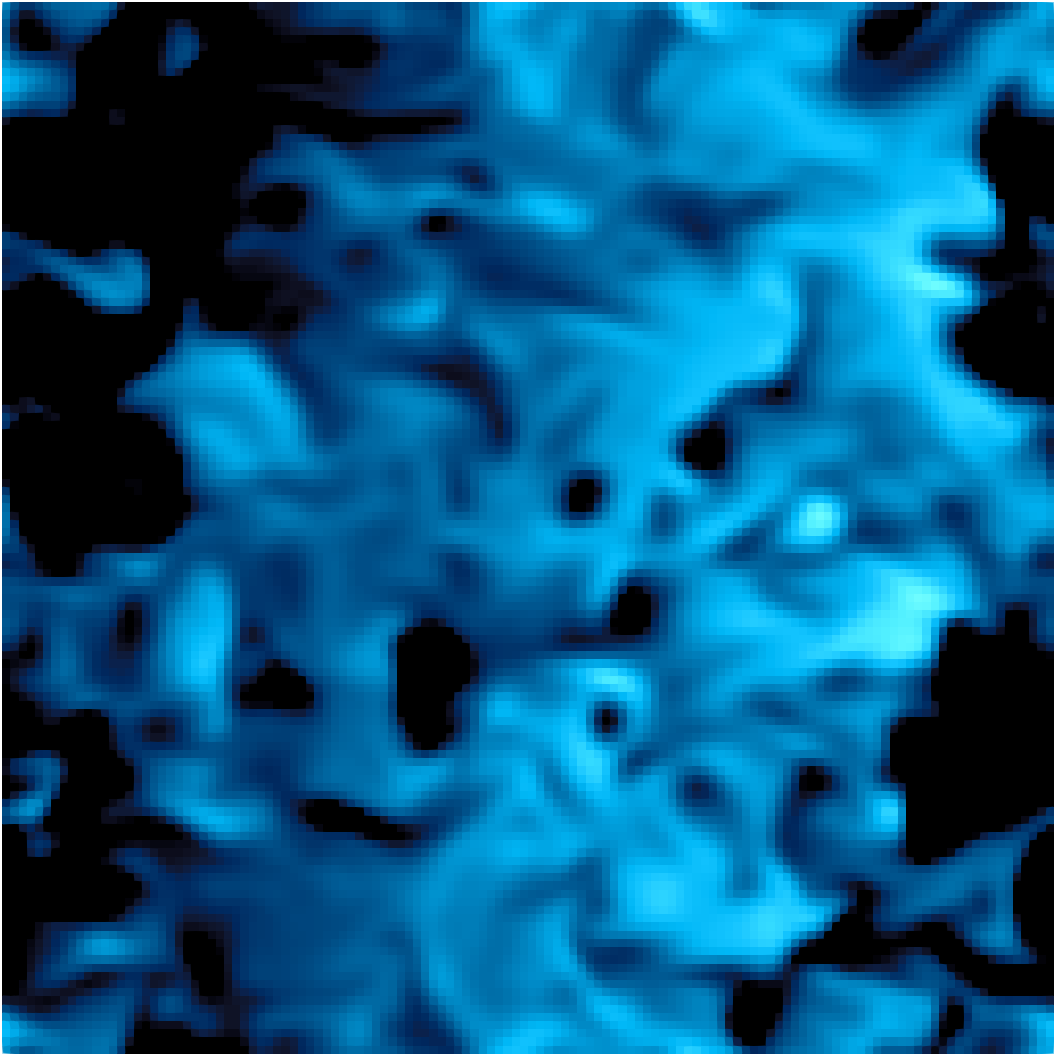}
     \includegraphics[width=0.22\textwidth]{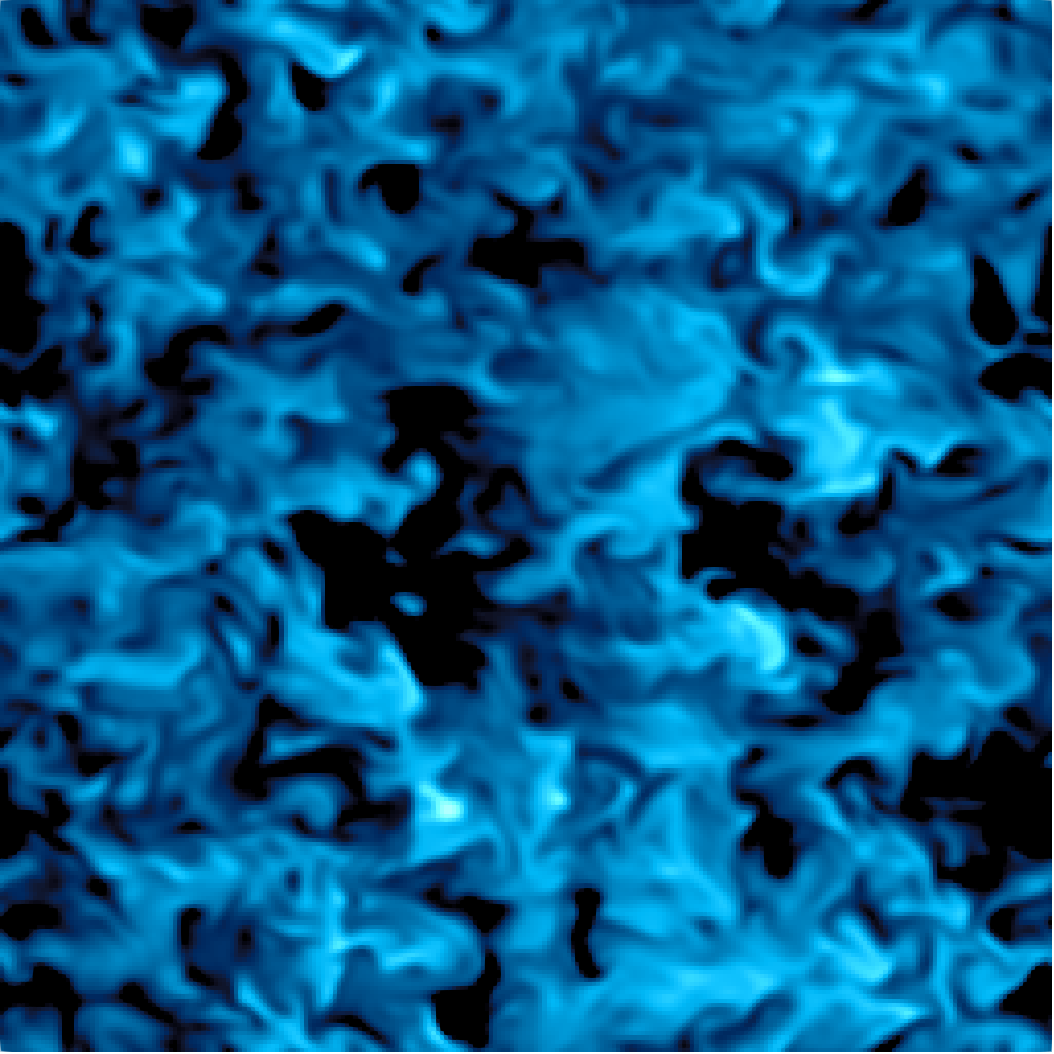}
     \includegraphics[width=0.22\textwidth]{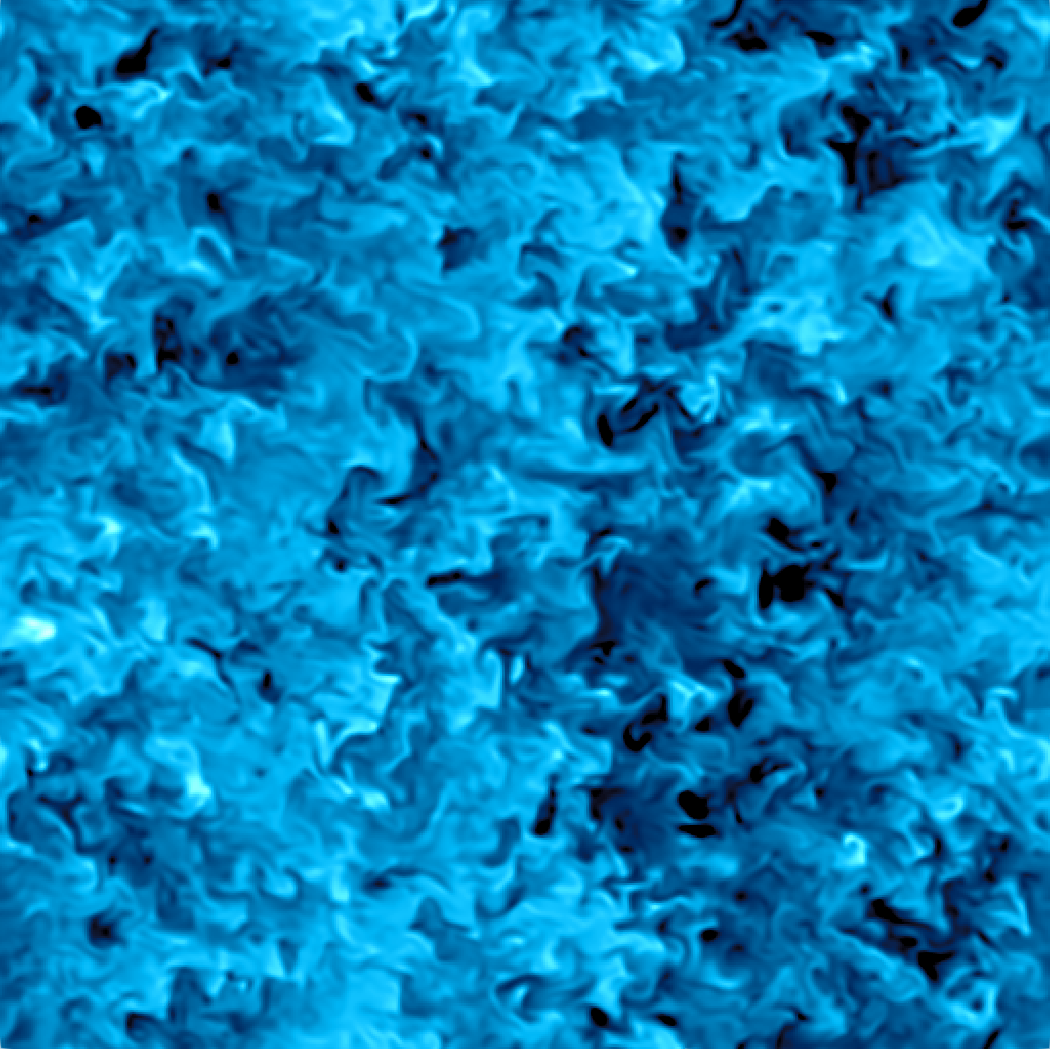}
     \includegraphics[width=0.04\textwidth]{./figs/bar-n512-z100.png}
     \hspace{-4em}\\
     \footnotesize{AMR-Wind, $z$= 100 m, $\Delta x$= 3.12 m, 1.56 m, 0.78 m ($n=128^3, 256^3, 512^3$)} \\
   \caption{\label{fig:512}
    NekRS(HPF, SMG) and AMR-Wind at three grid-refinement 
    levels for potential temperature at time 6 h.}
  \end{center}
\end{figure*}

\section{Simulations}
\label{sec:sim}
 
\begin{figure*}[t]
  \begin{center}
     \includegraphics[width=0.35\textwidth]{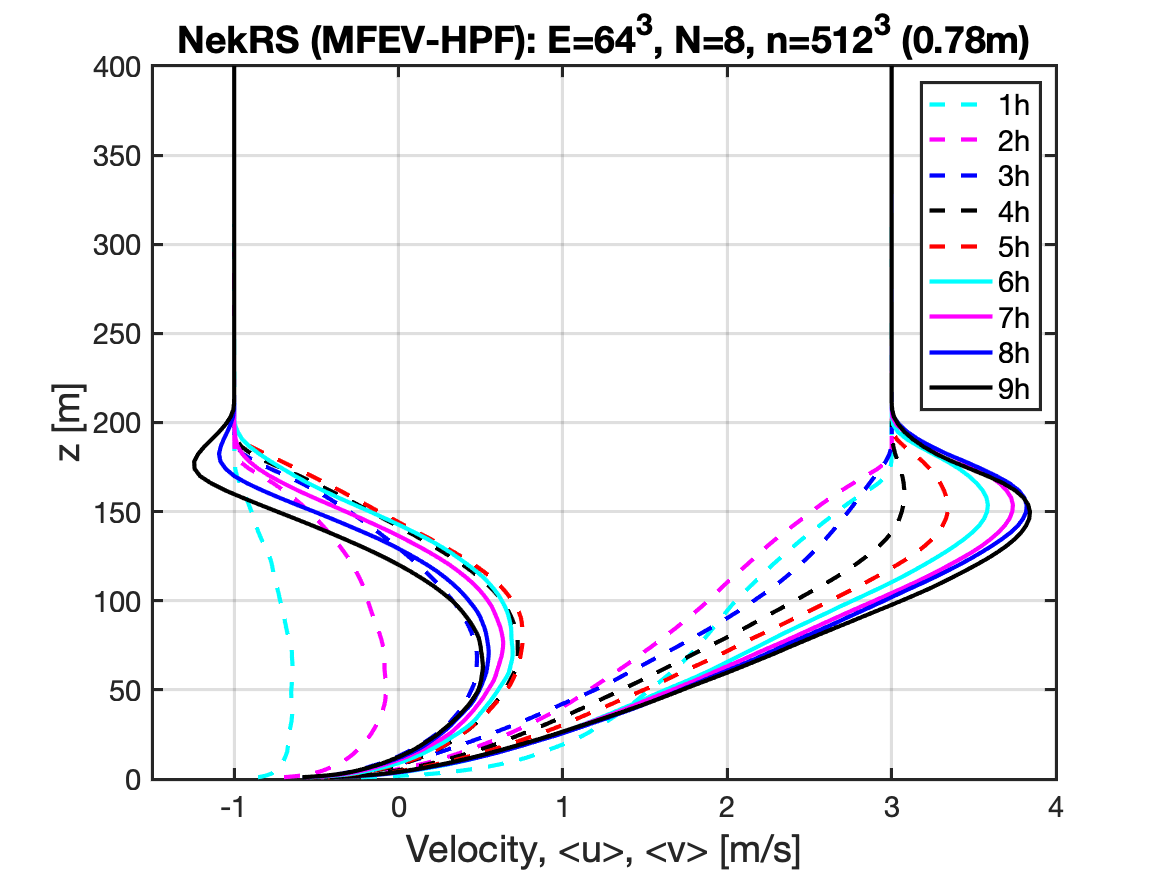}
     \hspace{-2em}
     \includegraphics[width=0.35\textwidth]{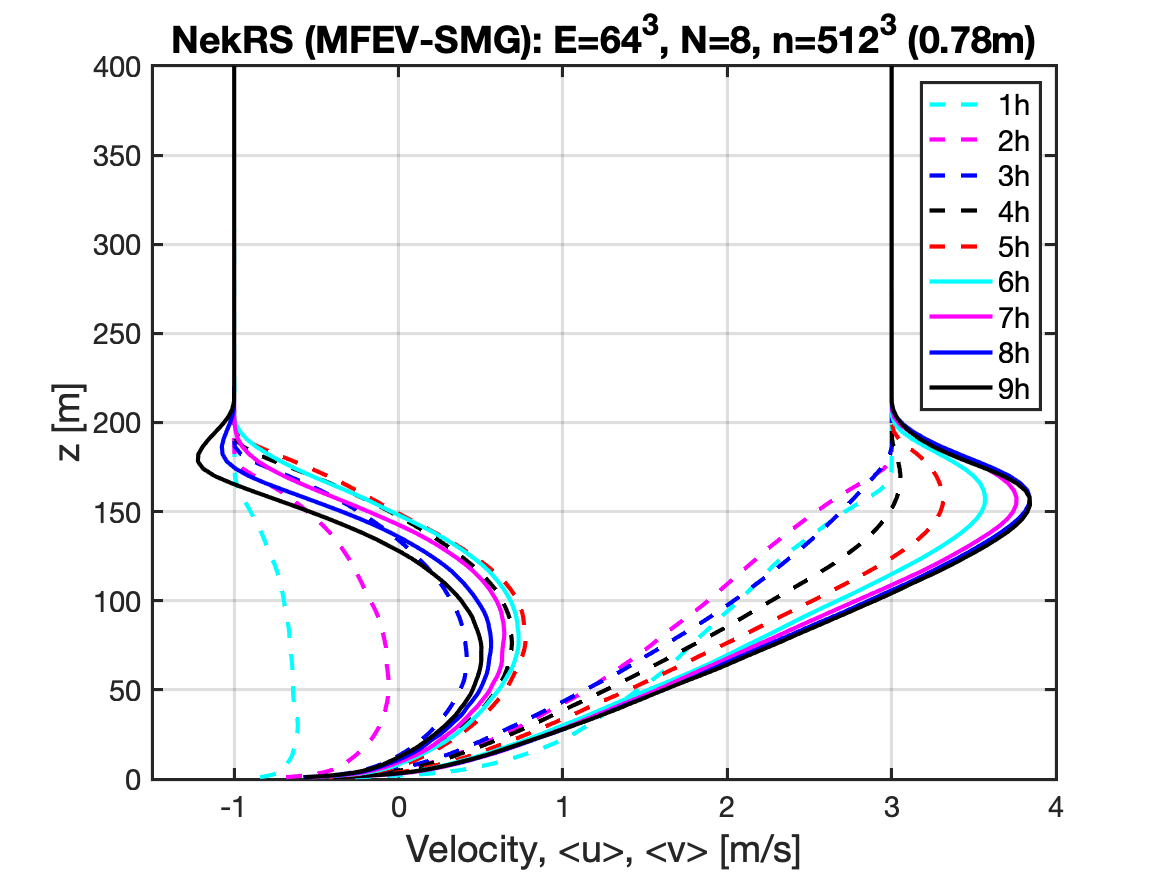}
     \hspace{-2em}
     \includegraphics[width=0.35\textwidth]{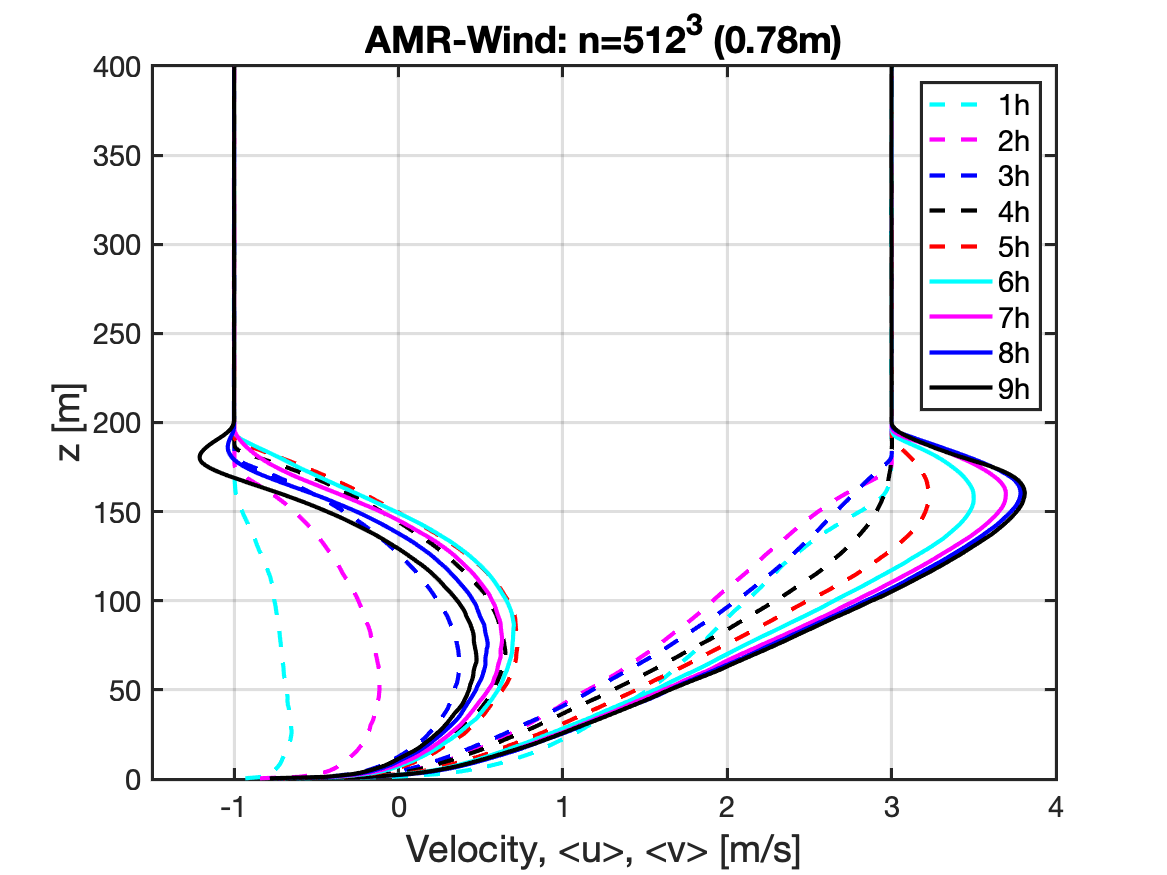}
     \\
     \hspace{-2em}
     \includegraphics[width=0.35\textwidth]{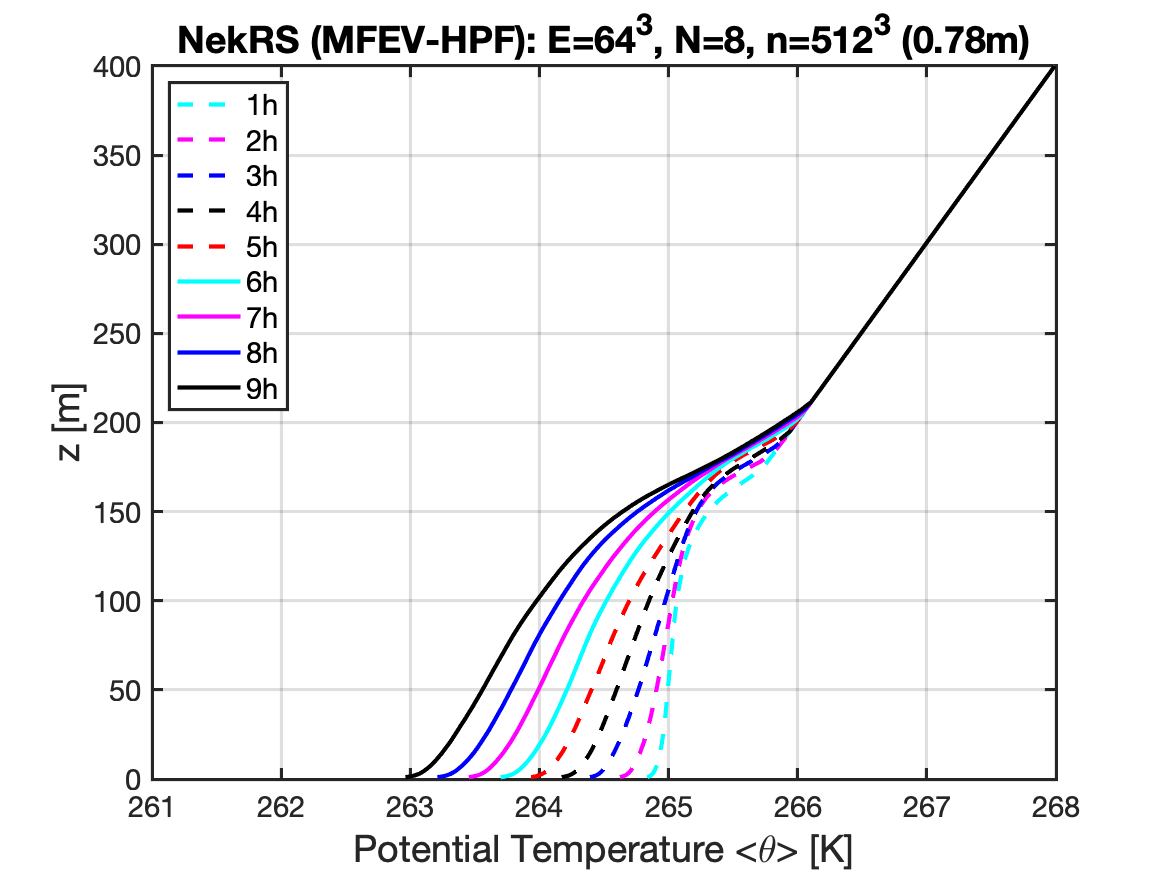}
     \hspace{-2em}
     \includegraphics[width=0.35\textwidth]{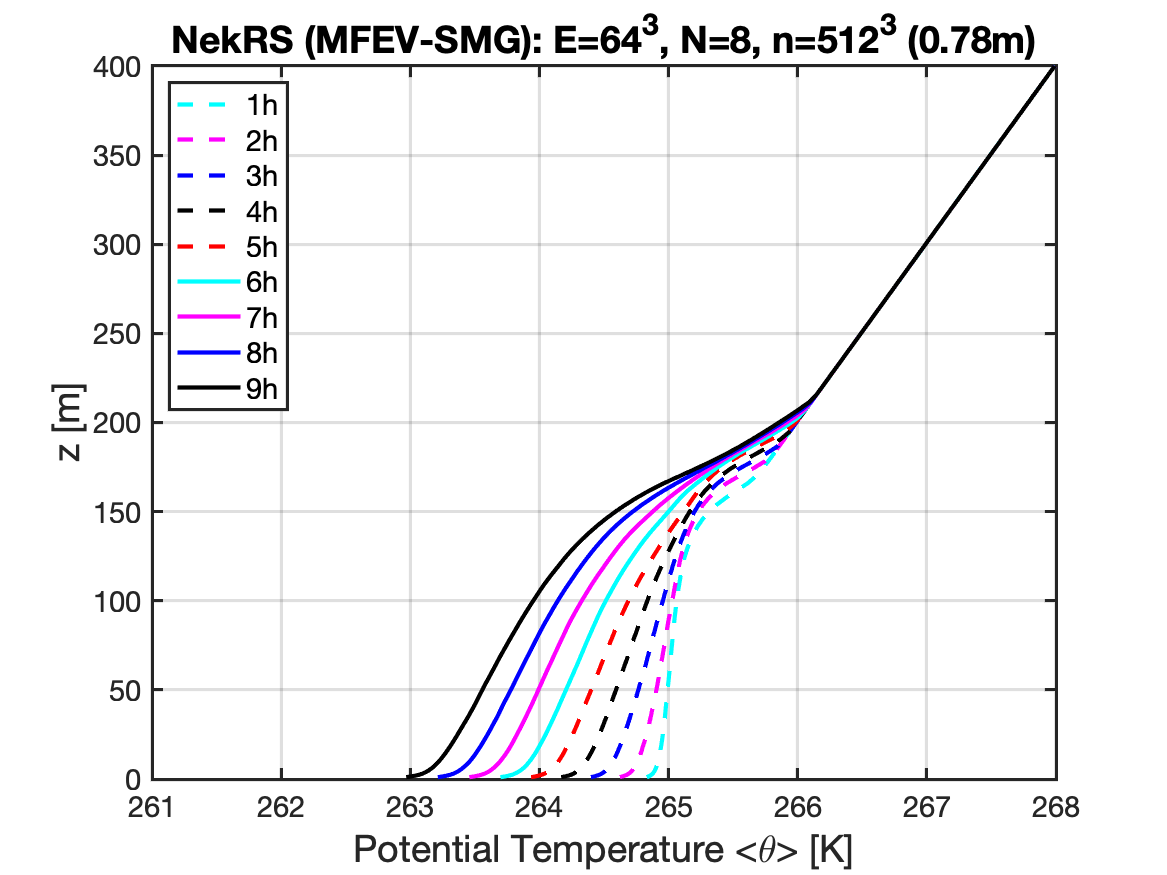}
     \hspace{-2em}
     \includegraphics[width=0.35\textwidth]{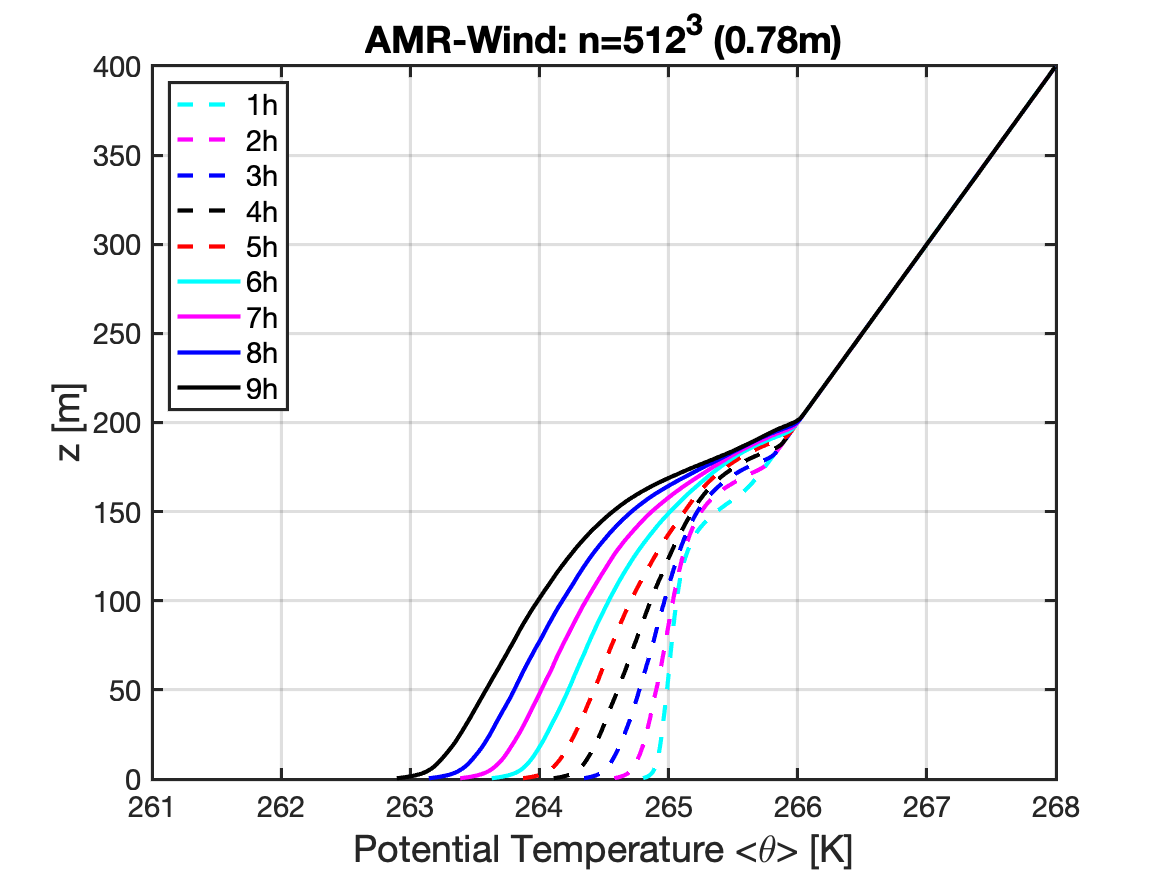}
     \hspace{-2em}
   \caption{\label{fig:mean}NekRS (HPF, SMG) and AMR-Wind: 
     velocity magnitude and potential temperature at each hour, 1 h, 2 h,...,9 h.
     }
  \end{center}
\end{figure*}

 We consider a stable ABL in which the ground temperature (at $z=0$) is cooler
 than the air temperature and where the ground temperature continues to cool over the 
 duration of the simulation.  Here we give full details of the numerical setup.
 The simulation domain is 
$\Omega = L_x \times L_y \times L_z =$ 400 m $\times$ 400 m $\times$ 400 m,  
with $x$ the streamwise direction, $y$ the spanwise direction, and
$z$ the vertical direction.
 Simulations are initialized (at $t=0$) with constant velocity in the
streamwise direction equal to geostrophic wind speed of $U=8$ m/s. The initial
potential temperature is 265 K in $0\le z\le 100$ m and linearly increased at
a rate of 0.01 K/m in 100 m $\le z\le 400$ m.  The reference potential
temperature is 263.5 K.  The Reynolds number is $Re= UL_b/\nu$, 
where $L_b=100$ m is the thickness of the initial
thermal boundary layer and $\nu$ is the molecular viscosity.
 An initial perturbation is added to the temperature with an amplitude of 0.1 K
 on the potential temperature field for $0 \le z\le 50$ m. 

Periodic boundary conditions (BCs) are used in the streamwise and spanwise directions.
At the top boundary, ($z=400$ m), a stress-free, rigid lid is applied for momentum, and
the heat flux for the energy equation is set consistent with the 0.01 K/m temperature
gradient initially prescribed in the upper region of the flow.
At the bottom boundary, we perform simulations with impenetrable traction BCs
for the velocity where the specified shear stress comes from Monin--Obukhov
similarity theory~\cite{monin1954}.  For the energy equation, a heat flux is
applied that is derived from the same theory and a specified potential
temperature difference between the flow at a height, $z_1$, and the surface.
The surface temperature is from the GABLS specification following the rule
$\theta_b(t) = 265 - 0.25 t$, where $t$ is in hours.  Because the boundary
conditions are periodic (lateral), or the mass flow rate through the boundaries
is zero (top and bottom),  pressure boundary conditions are not needed.

 In Nek5000/RS, the implementation of the traction BCs for the horizontal
velocity components is performed in the context of the log-law for which we
follow the approach of ~\cite{Grotjans_Menter1998} and~\cite{Kuzmin2007},
which is appropriate for finite element methods based on a weighted residual
formulation.
 The traction BCs imposed on the tangential velocity are based on the
horizontally averaged slip velocity that develops at the boundary and the law
of the wall and is effected through the use of the mean-field eddy viscosity
model of~\cite{sullivan94}. The approach originally used
by~\cite{SCHUMANN1975376} is used to convert the horizontally averaged traction
to local values based on the local slip velocity in each of the horizontal
directions.

 In AMR-Wind, the periodic BCs in the $x$ and $y$ directions and
 the slip boundary on the top wall are applied. On the bottom wall,
 Dirichlet BCs in the normal direction and inhomogeneous Neumann BCs in the $x$ and $y$ 
 directions are applied. The inhomoegenous Neumann  BC is set 
 using the expression for $\tau$, the total wall shear stress, and $q$, the total
 wall heat flux, in~(\ref{tau}).
 The stresses are specified at the terrain boundary following 
 Moeng~\cite{Moeng:1984}. The wall stress vector is defined as 
 \begin{eqnarray}
  \label{tau}
  \tau_{i3} = \frac{\bar{u_i}s + \bar{s} \left( u_i - \bar{u_i}\right)}{\bar{s}^2} u_\tau^2,
 \end{eqnarray}
 where $u_i$ is the velocity at the first cell height, $s$ is the wind speed 
 $s = \sqrt{u_1^2+ u_2^2}$, and $u_\tau$ is the friction velocity computed by using the 
 Monin--Obukhov similarity law~\cite{monin1954,Etling1996}. The overbar indicates a horizontal 
 plane average at the first cell height. The heat flux is defined as
 \begin{eqnarray}
  q =  \left [\left( \theta - \bar{\theta} \right) \bar{s} + 
       \left( \bar{\theta} - \theta_w\right) s \right ] \frac{u_\tau \kappa}{\bar{s} \phi_h},
 \end{eqnarray}
 where $\theta$ is the temperature,  $\theta_w$ is the wall temperature, $\kappa$ is
 the von Karman constant, and $\phi_h$ comes from the Monin--Obukhov similarity law.

The range of scales in these simulations is evident in Fig.~\ref{fig:512},
which shows variations in potential temperature on a horizontal $x$-$y$ slice
at the height $z=100$ m. for resolutions $\dx= 3.12$ m, $1.56$ m, and $0.78$ m, 
respectively from left to right, for the differing codes/models.  For NekRS,
$\dx$ represents the average grid spacing given by 400 m/$(E_*N)$, where $E_*$
is the number of elements in the $x$-, $y$-, or $z$-direction and $N$ is the
local polynomial order.  The number of elements is $E=E_*^3=16^3$, $32^3$, and
$64^3$ for the stated resolutions.  The top row in Figure~\ref{fig:512} shows
NekRS results using MFEV and HPF; the middle row shows NekRS results using MFEV
and Smagorinsky, as described in Section~\ref{sec:codes}; and the bottom row
shows AMR-Wind results.  At a height of $z=100$ m, the temperature variations
around the horizontally averaged value are small, between 264.40 K and 264.80 K.
One can see that as the grid scale is decreased from $\Delta x= 3.12$ m 
to $0.78$ m, both codes capture increasingly finer scales.   We remark that
direct numerical simulation at the given Reynolds number, $Re=5\times 10^7$,
would require $\approx 10^{15}$ grid points, which is two orders of magnitude
beyond current state of the art simulations of $n=18000^3$ for isotropic
turbulence \cite{pkyeung19}.  The importance of the SGS model is that it
potentially allows one to account for the effects of small-scale motions
without needing to resolve all of them.

Figure~\ref{fig:mean} shows profiles of the horizontally averaged streamwise,
$\left< u \right>$, and spanwise, $\left< v \right>$, wind velocities (top) and
potential temperature, $\left< \theta \right>$, (bottom) at 1-hour time
intervals between 1 h and 10 h for a mesh resolution of $n=512^3$ ($\dx= 0.78$ m)
for each code.  The left figures show NekRS results using MFEV and HPF, the
center figures show NekRS using MFEV and Smagorinsky, and the right figures
show the results for AMR-Wind.  As can be observed, the time evolution of the
mean velocity and temperature profiles obtained from the two codes agree 
well despite using very different numerical methods and subgrid-scale models.
The agreement between AMR-Wind and NekRS improves when using MFEV and
Smagorinsky (i.e., between the center and right figures).  Specifically, the
height of the low-level jet peak velocity during quasi-steady evolution in the
GABLS problem (after approximately 7 h) is between $150$ and $160$ m, and its
maximum value is between $9.5$ and $9.7$ m/s. An outgrowith of these comparative
simulations has been a concerted effort to carefully validate and cross-check
the SGS models.  The improvement in the NekRS SGS model, realized by moving
away from the HPF model to using the Smagorinsky model for the isotropic stress
term, is a direct outcome of this collaborative effort.

 \section{Performance}

 \begin{table*} 
  \footnotesize
  \begin{center}
  \begin{tabular}{|c|c|c|c|}
  \hline
  \multicolumn{4}{|c|}{{\bf Strong Scaling Test Sets}}\\
  \hline
    Domain size &  Grid Points ($n$)&  $\dx$ (m) & $\dt$ (s) \\
  \hline
  $[$400 m$]^3$ & 512 $\times$ 512 $\times$ 512  & 0.78 & .062500  \\
  $[$400 m$]^3$ &1024 $\times$ 1024$\times$ 1024 & 0.39 & .031250  \\
  $[$400 m$]^3$ &2048 $\times$ 2048$\times$ 2048 & 0.19 & .015625 \\
     -                                 & -                          &   -  &    -    \\
  \hline
  \end{tabular}
  \begin{tabular}{|c|c|c|c|}
  \hline
  \multicolumn{4}{|c|}{{\bf Weak Scaling Test Sets}}\\
  \hline
    Domain size &  Grid Points ($n$)&  $\dx$ (m) & $\dt$ (s) \\
  \hline
      400 m $\times$  400 m $\times$ 400 m & 512  $\times$ 512  $\times $512 & 0.78 &.0625   \\
      800 m $\times$  800 m $\times$ 400 m & 1024 $\times$ 1024 $\times $512 & 0.78 &.0625   \\
     1600 m $\times$ 1600 m $\times$ 400 m & 2048 $\times$ 2048 $\times $512 & 0.78 &.0625   \\
     3200 m $\times$ 3200 m $\times$ 400 m & 4096 $\times$ 4096 $\times $512 & 0.78 &.0625   \\
  \hline
  \end{tabular}
\end{center}
\caption{\label{cases}Problem setup for strong and weak scaling studies. }
\end{table*}

Here we compare performance and tuning for the two codes.  For each case, the
codes use the same spatial resolution, $\dx$, and timestep size, $\dt$.  Each
code uses iteration tolerances of $10^{-4}$ and $10^{-6}$ for the respective
2-norm residuals of the pressure-Poisson and velocity-Helmholtz problems.  For
purposes of timings, we use the solution at 6 hours as an initial condition in
each case in order to ensure that performance studies are done over a timeframe
in which the solutions have a representative turbulent flow.  
Table~\ref{cases} provides a summary of the test parameters, in physical units,
that are used for the strong- and weak-scaling studies.  The spectral element
cases use 8$th$-order polynomial basis ($N=8$) with a number of gridpoints
given by $n=EN^3$. 
For these cases we take $\dx$ to be the average grid spacing in
each direction (i.e., 400~m / $n^{\frac{1}{3}}$).  For the weak-scale study,
the domain height is fixed at 400~m while the dimensions are increased in the
$x$ and $y$ directions as $n$ is increased.  In order to avoid initial transient
behavior, the average (wall) time per step, $t_{step}$, in seconds is measured
over steps 101--200.

%

\subsection{Performance Tuning and Profiling} \label{sec:tune}
We begin with performance optimization, profiling analysis, and
CPU versus GPU comparisons.


 \begin{table*}[t]
  \footnotesize
  \begin{center}
  \begin{tabular}{|l||r|r||r|r||r|r|}
  \hline
  \multicolumn{7}{|c|}{{\bf AMR-Wind: Performance progress with AMReX library updates} }\\
  \multicolumn{7}{|c|}{Last 200 steps averaged from 1000-step run on 8 nodes using $n=512^3$ }\\
  \hline
   \multicolumn{1}{|c||}{}   &
   \multicolumn{2}{|c||}{Old Version} &
   \multicolumn{2}{|c||}{Intermmediate Version} &
   \multicolumn{2}{|c|}{New Verision}\\
   \hline
                     &   (s)   & [\%] & (s)   & [\%] & (s)   & [\%] \\
    Wall time per timestep      &  3.3200e-01 & 100 & 2.4100e-01  & 100 & 2.2500e-01  & 100 \\
   \hline
        Advection                   &  2.9814e-02  &   8.98  &  2.6920e-02   & 11.17   &  2.8687e-02  & 12.75 \\
        MAC Projection              &  6.2582e-02  &  18.85  &  6.2660e-02   & 26.00   &  6.3135e-02  & 28.06 \\
        Pressure Solve              &  7.3671e-02  &  22.19  &  7.3481e-02   & 30.49   &  6.3180e-02  & 28.08 \\
        Velocity Solve              &  1.1401e-01  &  34.34  &  3.9669e-02   & 16.46   &  3.9307e-02  & 17.47 \\
        Scalar Solve                &  3.4827e-02  &  10.49  &  2.2148e-02   &  9.19   &  1.5930e-02  &  7.08 \\
        Fillpatch                   &  1.5538e-02  &   4.68  &  1.4990e-02   &  6.22   &  1.5030e-02  &  6.68 \\
  \hline
  \end{tabular}
\end{center}
\caption{\label{amr_op}AMR-Wind performance optimization.}
\end{table*}

   NekRS GPU performance tuning on Summit is demonstrated in detail
in~\cite{nekrs,gb21}.  The base libParanumal kernels have their origins 
in the work of Warburton and co-workers
\cite{libp,warburton2019,warburton2020,ceed_special_issue2}. 
   A key algorithmic component is the Chebyshev-accelerated Schwarz-based
$p$-multigrid for the pressure solve \cite{malachi2022a}, which is performed in
32-bit precision (e.g., as in \cite{fehn18b}) to reduce injection-bandwidth
pressure on the Summit network interface cards.  
  Communication for the nearest-neighbor communication (direct-stiffness
summation in the finite element or spectral element context \cite{dfm02})
is overlapped with computation whenever it proves to be effective,
which can yield as much as 10--15\% savings in NS applications.
At the strong-scale limit of $\approx 2$M points per GPU, there are
enough points interior to each rank's subdomain to balance out the
communication overhead for the gather-scatter exchanges, at least at the
fine-mesh level evaluations.  For the coarser $p$-multigrid levels, it is not
always the case that one can cover the communication with work.  When
initializing the communication kernels for each level of the $p$-multigrid
solver, NekRS selects the fastest of several available communication strategies
(e.g., overlapping, pack-on device or host, GPU direct or via the host), which
are determined by timing tests during runtime setup.  These unit tests also
report the observed messaging bandwidth and thus provide insight into possible
anomalous system behavior, which is useful when porting relatively new code to
relatively new and unknown HPC platforms.  For example, from existing logfiles,
we were able to compare the observed bandwith for several gather-scatter
exchanges on NERSC's Perlmutter platform before and after a network update from
Cray's Slingshot 10 to Slingshot 11, as shown below. 
\tiny
\footnotesize
\hspace*{.04in}
\begin{minipage}{3.5in}
\begin{verbatim}

SS10:
pw+device MPI: 7.37e-05s / bi-bw:  54.5GB/s/rank
pw+device MPI: 5.16e-05s / bi-bw: 100.2GB/s/rank
pw+device MPI: 3.84e-05s / bi-bw:  33.6GB/s/rank
pw+host   MPI: 2.46e-05s / bi-bw:   3.6GB/s/rank

SS11:
pw+device MPI: 4.38e-05s / bi-bw:  91.8GB/s/rank
pw+device MPI: 3.47e-05s / bi-bw: 148.8GB/s/rank
pw+device MPI: 2.74e-05s / bi-bw:  47.2GB/s/rank
pw+host   MPI: 1.66e-05s / bi-bw:   5.4GB/s/rank

\end{verbatim}
\end{minipage}
\normalsize

\noindent
Here, SS10 indicates Slingshot 10, and SS11 indicates Slingshot 11, which
shows about a 1.5$\times$ improvement over SS10.  The listings also show which
communication mode was used. We see that {\tt pw+device}, which stands for
pairwise device-to-device exchange (i.e., via GPU-direct) is used in most
instances.  The {\tt pw+host}, which indicates the use of pairwise exchanges
via the host, is used only in the case of many short messages, which is
typically the scenario at the coarsest levels of the $p$-multigrid solver.

%

 \begin{table*}[t]
 \footnotesize
   \begin{center} \begin{tabular}{|r|r|r|r|l|l|}
   \hline
   \multicolumn{6}{|c|}{{\bf Nsight-Compute Profiling: CUDA Kernel Statistics}}\\
   \multicolumn{6}{|c|}{11 nodes (66 GPUs), $n/P=2.03M$, $n=512^3$, 2000 steps}\\
   \hline
   \hline
   \multicolumn{6}{|c|}{{\bf NekRS}}\\
   \hline
   Time [\%]  & Total Time (ms) & Instances  & Average ($\mu$s) &    Name & Remark \\
   \hline  
     7.2   &    1438.604 &      3327  &    432.402  &  subCycleStrongCubatureVolumeHex3D    & dealiased vel. adv.    \\  
     6.6   &    1320.502 &      8002  &    165.021  &  gatherScatterMany\_doubleAdd         & FP64 local gather-scatter    \\
     5.1   &    1017.806 &      8144  &    124.976  &  packBuf\_doubleAdd                   & FP64 packing for gs    \\
     5.0   &    1009.247 &      3533  &    285.663  &  subCycleStrongCubatureVolumeHex3D    & dealias scalar adv.    \\ 
     4.7   &     935.298 &       251  &   3726.289  &  scatterMany\_double                  & FP64 scatter    \\
     4.4   &     879.495 &      8144  &    107.993  &  unpackBuf\_doubleAdd                 & FP64 gather    \\
     3.3   &     667.776 &      3192  &    209.203  &  subCycleRKUpdate                     & RK4 vector update    \\
     2.9   &     577.150 &       850  &    679.000  &  ellipticStressPartialAxCoeffHex3D    & viscous op. eval.    \\
     2.4   &     480.331 &      2788  &    172.285  &  ellipticPartialAxHex3D               & pressure op. eval.    \\
   \hline
   \hline
   \multicolumn{6}{|c|}{{\bf AMR-Wind}}\\
   \hline
   Time [\%]  & Total Time (ms) & Instances  & Average ($\mu$s)   &  Name & Remark \\
   \hline
    15.0    &  1890.142  &   142823 & 13.234  &  fab\_to\_fab               &   array box local copy\\
     9.9    &  1256.128  &   12800  & 98.148  &  MLNodeLaplacian::Fsmooth   & multigrid smoother  \\
     6.9    &   873.629  &   24800  & 35.227  &  amrex::Copy                &  multiple array box parallel copy \\
     5.9    &   738.200  &   5600   & 131.821 &  MLABecLaplacian::Fapply    & Laplacian op. eval.  \\
     4.5    &   564.742  &   43200  & 13.072  &  MLPoisson::Fsmooth         & multigrid smoother  \\
     3.5    &   438.271  &   6800   & 64.451  &  MultiFab::LinComb          & vector-vector addition  \\
     3.1    &   394.024  &   3200   & 123.132 &  MLABecLaplacian::normalize & normalize solution  \\
     3.0    &   384.391  &   800    & 480.488 &  godunov::compute\_fluxes   & advection momentum \\
     2.9    &   359.910  &   800    & 449.887 &  godunov::compute\_fluxes   & advection  scalar\\
     2.7    &   344.575  &   11800  & 29.201  &  MultiFab::Xpay             &  vector-vector addition \\
     2.4    &   303.2    &   29850  & 11.085  &  FabArray::setVal           & set value of array box \\
   \hline
   \end{tabular}
   \end{center}
   \caption{\label{prof1}
    CUDA kernel statistics from
    {NVIDIA$\circledR$  Nsight\textsuperscript{TM}} profiler
    using  {\tt nsys profile --stats=true -t nvtx,cuda}.}
 \end{table*}

Over the course of the collaboration, AMR-Wind realized a 1.4$\times$ speedup
with some improvements derived through AMReX library updates.  The performance
progress is demonstrated in Table~\ref{amr_op}, where
   the rows present a timing breakdown of a typical flow time step.
Advection involves predicting and forming the advection term using Godunov PPM
WENO. MAC projection is a Poisson equation linear solve with a 7-point stencil
that ensures that the face velocities are divergence free. The pressure solve is
a Poisson equation linear solve with a 27-point stencil that approximately
corrects the cell velocity to be divergence free at the end of the time step.
Velocity and scalar solve are Helmholtz equations with a 7-point stencil, and
Fillpatch performs all communication within and across processors
outside of the linear solver communication.
  In the table, the {\em old version} is AMR-Wind using AMReX from April 2021.
The {\em intermediate version} is the same source code 
but with improvements to the linear solver settings. 
In particular, the components of the momentum equations are solved separately instead of 
as a coupled tensor solve.
The velocity and scalar (temperature) linear systems are solved by using bi-conjugate
gradient iteration instead of a full geometric multigrid approach.
In Table~\ref{amr_op} we see that these optimizations reduce the velocity
solve time by almost 3$\times$ (.114 s to .039 s) and the scalar solve
time by 1.5$\times$ (.035 s to .022 s).
  The {\em new version} is AMR-Wind based on AMReX from 2022 with the same 
improved linear system settings.  Here the scalar solve improves by another
factor of 1.5. and the pressure solve is reduced from .0073 s to .0063 s
per step.

\begin{figure*}[!h]
   \begin{center}
    \includegraphics[width=0.44\textwidth]{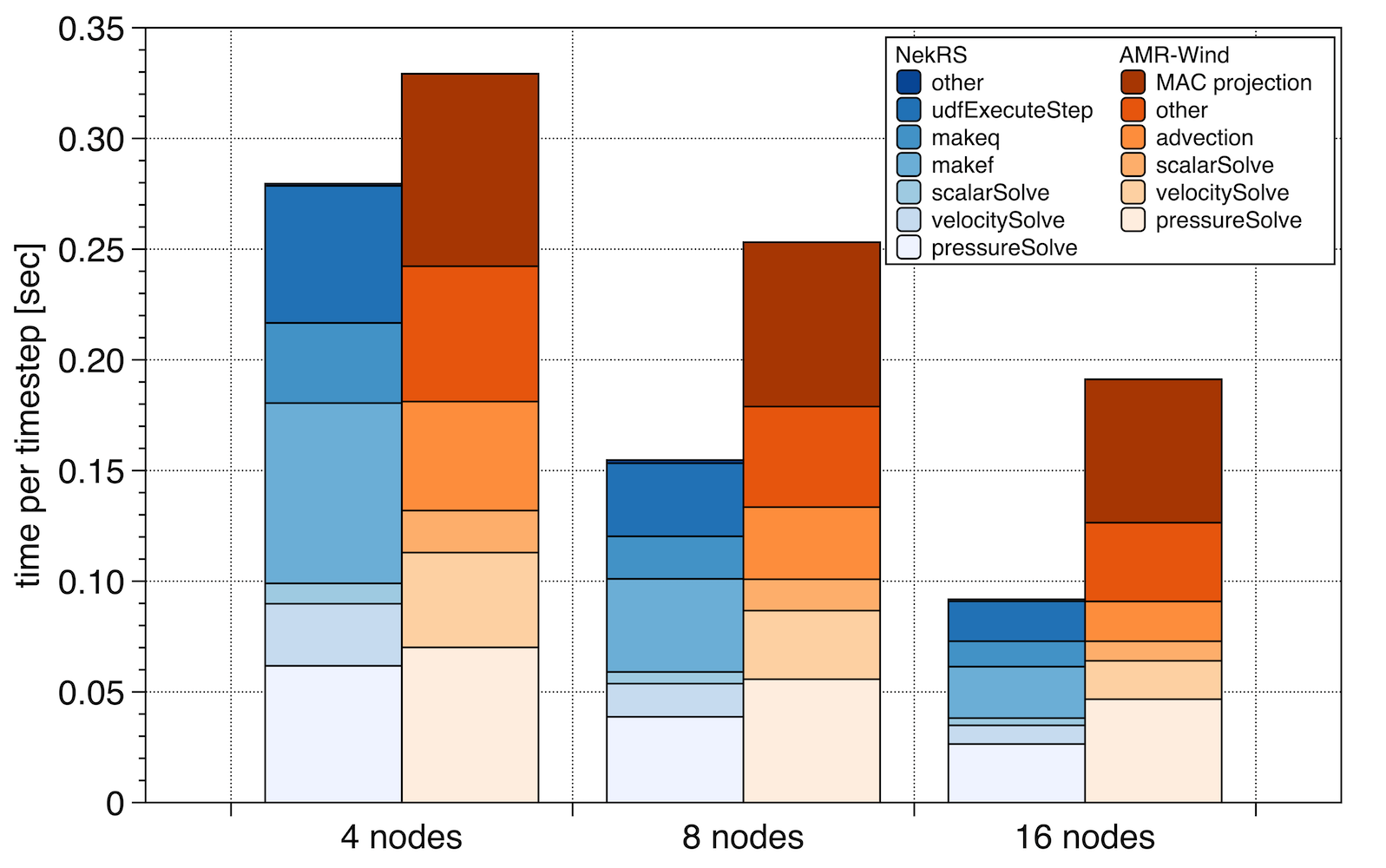}
    \includegraphics[width=0.44\textwidth]{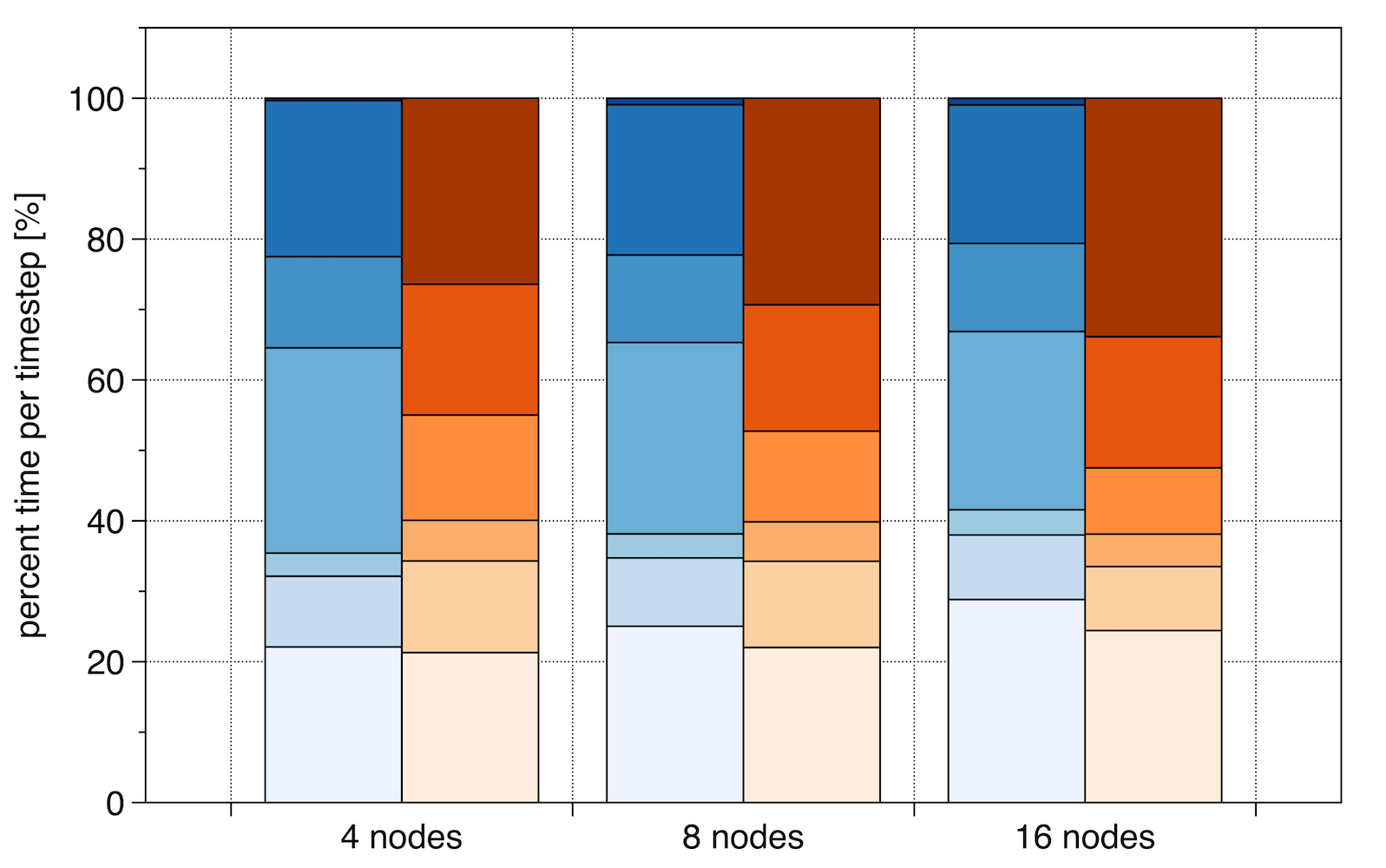}
    \\
    \includegraphics[width=0.44\textwidth]{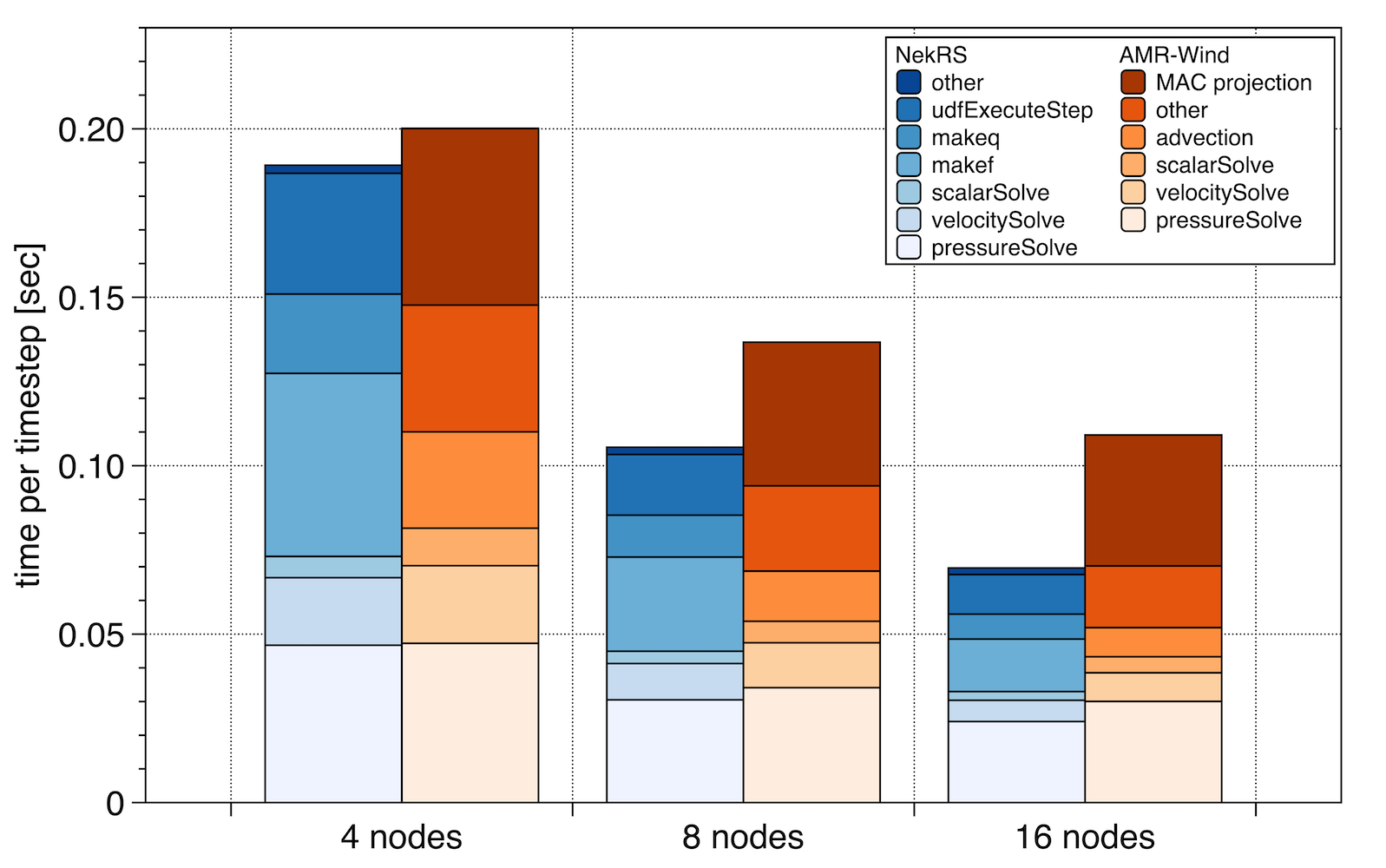}
    \includegraphics[width=0.44\textwidth]{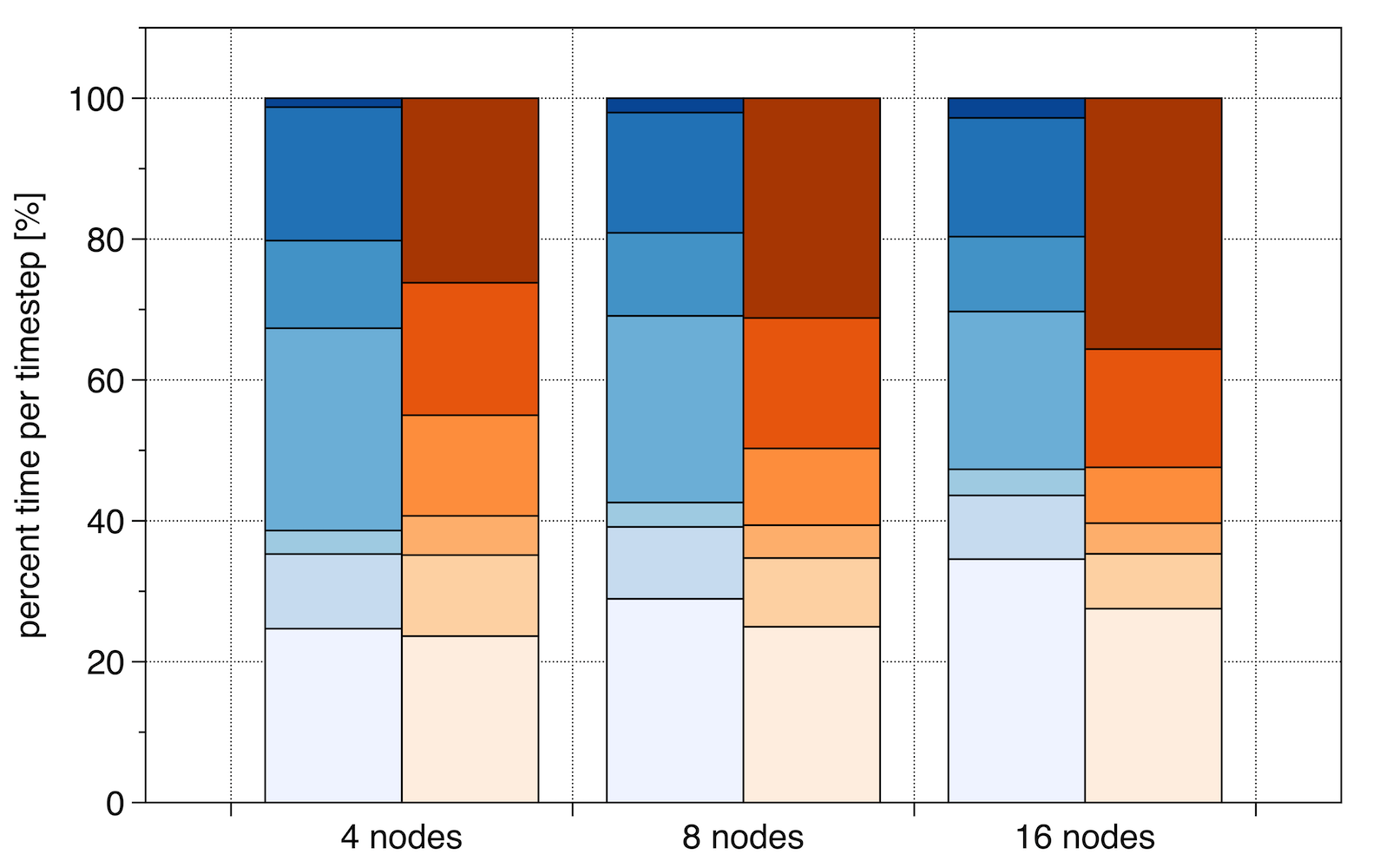}
  \caption{\label{fig:cost}NekRS vs AMR-Wind GPU cost breakdown on Summit (top) 
            and Crusher (bottom), using $n=512^3$ and 2000 steps.}
  \end{center}
\end{figure*}




For AMR-Wind, Table~\ref{prof1} more clearly indicates the elliptic solves as
leading cost contributors.   This cost is also reflected in
Fig.~\ref{fig:cost}, where the two largest contributors to run time are the
pressure solve and the MAC projection onto a divergence-free space.  In fact,
these plots show that the requirement of two Poisson-like solves for AMR-Wind
is the principal cause for discrepancy in run-time between the two codes.
MAC projection is solved using geometric multigrid. While not necessary, it does
provide more robustness and increases the stability of the scheme to CFL=2.
  If it did not require the MAC step, AMR-Wind would be faster on 4 nodes than
  NekRS.
pressureSolve is a Poisson solve that is used at the end of the timestep to
form an approximate divergence-free velocity at the cell center; it is a
node-based 27 point stencil, and the linear system is solved using geometric
multigrid.  scalarSolve and velocitySolve are both Helmholtz solves that are
cell-based 7 point stencils, the scalarSolve advances in time the potential
Temperature equation and velocitySolve is three separate solves to advance each
of the momentum equations in time. BiCG is used to solve all of the linear
Helmholtz subproblems.  The advection terms in the governing equations are
discretized using a Godunov WENO-Z scheme to provide these terms on the cell
faces at time $t^{n+\frac{1}{2}}$.  Other function calls comprise source
term calculations, boundary conditions, planar averaging, communication
(excluding linear solve communication), linear solve setup, and copying
solution arrays.

For AMR-Wind on both Summit and Crusher, the time per step, $t_{step}$,
decreases with increasing node count as each component of the timestep takes
less time.  Both Poisson solves, however, take a higher percentage of the time
step as $P$ is increased, which reflects the communication-intensive nature of
the Poisson problem.  Overall, Crusher is providing better performance than
Summit. This is partly because there are more GPUs per node (8 versus 6) but
also because the mesh decomposition has better load balancing for AMR-Wind. A
problem size of $512^3$ is more easily partitioned by 8 GPUs/node versus 6
GPUs/node. With 16 Crusher nodes (128 GPUs) a time per timestep of
$t_{step}$=0.11 s is achieved with AMR-Wind.  Further scaling out with Summit
the lowest time per timestep was 0.128 s on 128 Summit nodes (768 GPUs), as
discussed below.

\begin{table}[t] \footnotesize \begin{center} 
\begin{tabular}{|l||l|r|r|r|r|r|} \multicolumn{7}{c}{{}} \\ \hline
Platform  & Kernel     & $N$  &  FP   &  GB/s    & GFLOPS &KV \\ \hline
Summit    & advSub (3) &  10  &  64   &   613    &  3773  & 7 \\
NVIDIA    & advSub (1) &  10  &  64   &  1137    &  2446  & 8 \\
V100      & Ax         &   8  &  64   &   844    &  1622  & 5 \\
($P$=24)  & Ax         &   8  &  64   &   900    &  1731  & 4 \\
          & Ax         &   8  &  64   &   901    &  1732  & 4 \\
          & Ax         &   8  &  32   &   859    &  3303  & 4 \\
          & Ax         &   4  &  64   &   832    &   975  & 5 \\
          & Ax         &   4  &  32   &   711    &  1667  & 6 \\
          & fdm        &  10  &  32   &   611    &  6210  & 3 \\
          & fdm        &   6  &  32   &   713    &  4422  & 3 \\ \hline
Crusher   & advSub (3) &  10  &  64   &   491    &  3018  & 11 \\
AMD       & advSub (1) &  10  &  64   &   868    &  1867  & 8 \\
MI250X    & Ax         &   8  &  64   &   662    &  1272  & 2 \\
($P$=32)  & Ax         &   8  &  64   &   736    &  1416  & 2 \\
          & Ax         &   8  &  64   &   736    &  1415  & 2 \\
          & Ax         &   8  &  32   &   742    &  2854  & 2 \\
          & Ax         &   4  &  64   &   708    &   830  & 6 \\
          & Ax         &   4  &  32   &   658    &  1543  & 0 \\
          & fdm        &  10  &  32   &   546    &  5551  & 4 \\
          & fdm        &   6  &  32   &   521    &  3234  & 4 \\ \hline
Summit    & Sustained  &      &  64   &          &   833  &   \\ \hline
Crusher   & Sustained  &      &  64   &          &   937  &   \\ \hline 
\end{tabular}
\end{center}
\caption{\label{tab:kernel}NekRS runtime benchmark results associated with Fig. \ref{fig:cost}.}
\end{table}

For NekRS, we start the GPU analysis with NVIDIA's profiling tools.
Table~\ref{prof1} summarizes the kernel-level metrics for the critical kernels,
which are identified with NVIDIA's Nsight Systems.  At this granularity, the
table indicates that the bulk of the time for NekRS is spent evaluating the
dealiased advection operator (subCycleStrongCubatureVolumeHex3D) both for the
velicity vectors and for the temperature.    Other leading consumers are the
gather-scatter operations.   Largely missing from this table for NekRS is the
time spent in the pressure preconditioner, which is separated across many
kernels for the various levels of $p$-multigrid.   

Each NekRS job tracks basic runtime statistics using a combination of MPI
Wtime and cudaDeviceSynchronize or CUDA events. These are output every 500 time
steps unless the user specifies otherwise.  From these, we collect aggregate
timing breakdowns, roughly following the physical substeps of advection,
pressure, and viscous- thermal-updates, plus tracking of known communication
bottlenecks such as the pMG coarse-grid solve for the pressure preconditioner.
Figure~\ref{fig:cost} shows the cost breakdown for this type of analysis over
node counts ranging from 4 to 16.  At lower node counts, the bulk of the
NekRS time is spent in the makef and makeq (advection) routines, which are respectively
responsible for setting up the right-hand-sides for the momemtum and energy
equations.  To allow a larger CFL, the ABL simulations use characteristics-based
timestepping, which involves solving a sequence of hyperbolic subproblems on
the interval $[t^{n-2},t^{n}]$ (one for each velocity component and one for
temperature) \cite{maparo90,patel18}. Each subproblem takes several substeps
using the dealiased advection operator, which performs quadrature on a $11
\times 11 \times 11$ grid in each element. These substeps are thus compute-intensive
but not communication intensive, so they scale relatively well.
velocitySolve and scalarSolve, which involve communication-free
diagonal preconditioning for conjugate gradient solution of (\ref{vel})--(\ref{temp}),
show similar scaling behavior.
As with AMR-Wind, we see clearly in Fig. \ref{fig:cost} that the pressure solve
not scale as well as the other components.


  We remark that Fig. \ref{fig:cost} indicates a significant amount of time
is spent in udfExecuteStep.  The majority of that cost results from the
recently adopted mean-field eddy viscosity model (\ref{eqn.tijdef}), which
requires several planar averages per time step and is currently implemented as
a user-defined function.  For these calculations, which have low pressure and
velocity iteration counts, the frequently called planar average utility has a
significant impact on runtime (about 20\%).  Planar averaging is typically a
post-processing operation that is not performed on every step, but clearly
it will need to be optimized in this LES application.

Table \ref{tab:kernel} gives a detailed breakdown of the per rank\footnote{One
rank corresponds to a single V100 on Summit or a single MI250X GCD on Crusher.}
kernel performance for NekRS in the 4-node case of Fig. \ref{fig:cost}.  The
kernels are the advection subroutine (advSub), with either 3 components
(velocity) or 1 (temperature); the Poisson/Helmholtz matrix-vector product (Ax)
for the elliptic solves; and the fast-diagonalization method (fdm) for the
Schwarz smoother \cite{lottes05}.  For each kernel, $N$ indicates the
polynomial order. To leading order, the amount of tensor-contraction
work for each operation scales as $CE_p(N+1)^4$ and the number of memory
references as $CE(N+1)^3$,
with $E_p$ the number of elements on each rank and $C\approx 12$--30 a
kernel-dependent constant.  FP indicates the working floating-point precision;
GB/s the sustained streaming bandwidth on the device; GFLOPS the number of
billions of floating-point operations per second sustained on the device for
that particular kernel; and KV the kernel version identified 
as the fastest entry in each runtime benchmark test.  The 32-bit precision
kernels are used in the lower levels of $p$-multigrid.  Because the Schwarz
smoother operates on an extended domain, fdm executes on data that is extended
to $N+2$ in each direction compared with its corresponding Ax operation.  The
advection operation is dealiased (i.e., integration is on a finer mesh than the
underlying velocity representation), so that kernel also operates on a
relatively large data set.  We see that the work-intensive (high-$N$) and
32-bit kernels achieve impressive floating-point performance, well in excess of
1 TFLOPS (incidentally, the speed of ASCI Red, the world's fastest computer
just 25 years ago).  NekRS also provides a conservative estimate of the overall
FP64 floating point rate per rank---here, close to 1 TFLOPS---which includes
the message-passing overhead.  (For this overall rate, each 32-bit operation is
counted as half a flop.)

\begin{figure*}[!h]
   \begin{center}
    \includegraphics[width=0.38\textwidth]{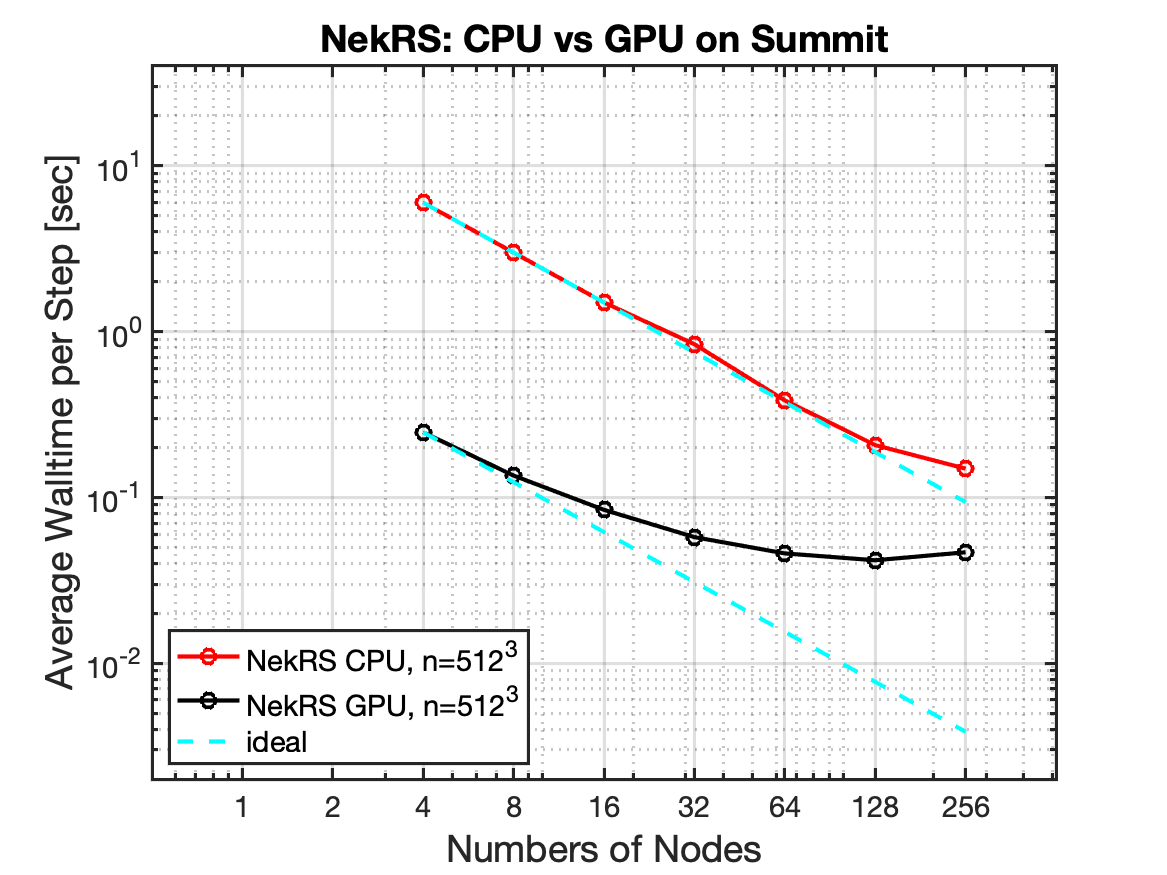}
    \hspace{-2em}
    \includegraphics[width=0.38\textwidth]{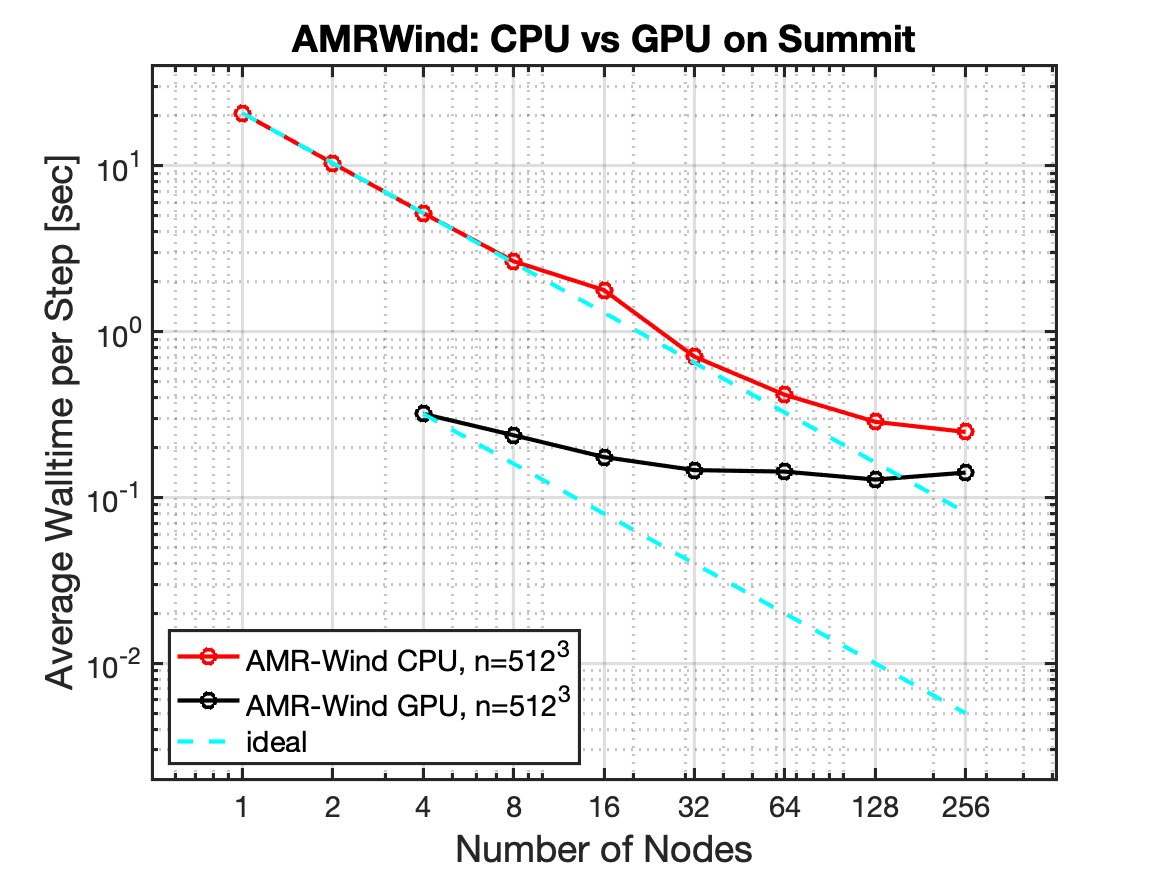}
  \caption{\label{cpu-gpu}NekRS and AMR-Wind: CPU vs. GPU performance on Summit:
            100 steps average from 200 step runs for $n=512^3$.}
  \end{center}
\end{figure*}

Figure~\ref{cpu-gpu} shows CPU and GPU strong-scaling performance for each code
on Summit.  The upper figures show standard time vs. node-count plots, which
clearly indicate that it is easier to strong-scale on the CPU.  
On Summit, however, that point is moot given that one needs 128 nodes using a
CPU-only configuration in order to get to the same time-per-step as using 4
nodes with 6 GPUs each (i.e., {\em a factor of 32 difference in required 
node-hours to do the same work}).

\subsection{Strong- and Weak-Scaling Performance}

\begin{figure*}[!h]
   \begin{center}
    \includegraphics[width=0.35\textwidth]{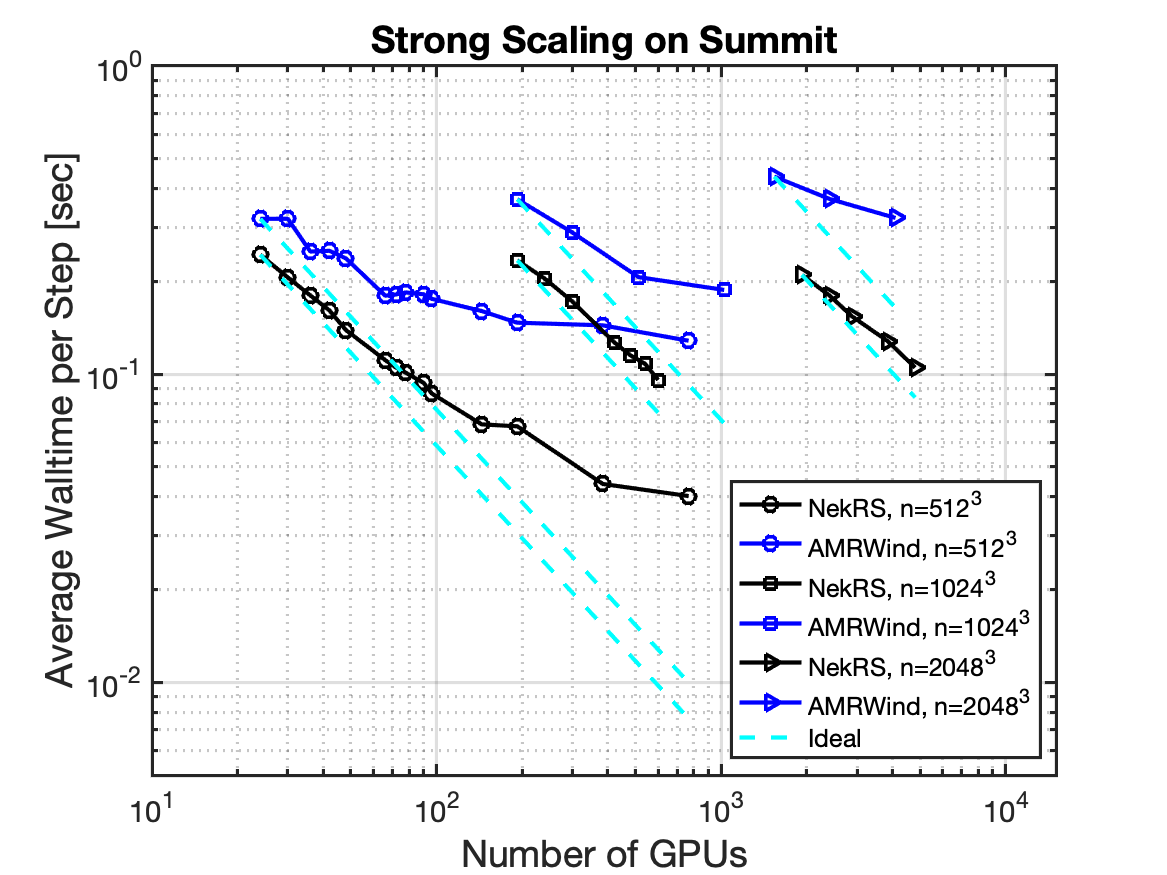}
    \hspace{-2em}
    \includegraphics[width=0.35\textwidth]{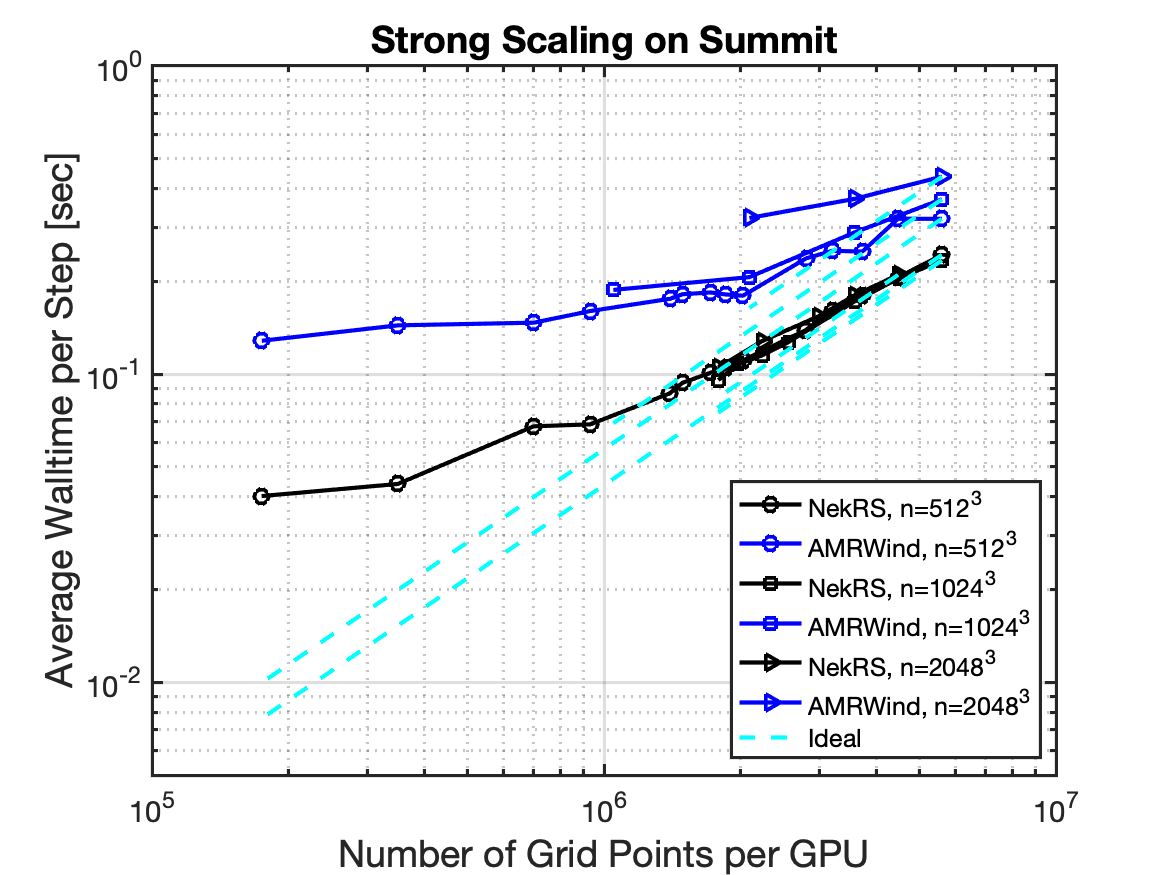}
    \hspace{-2em}
    \includegraphics[width=0.35\textwidth]{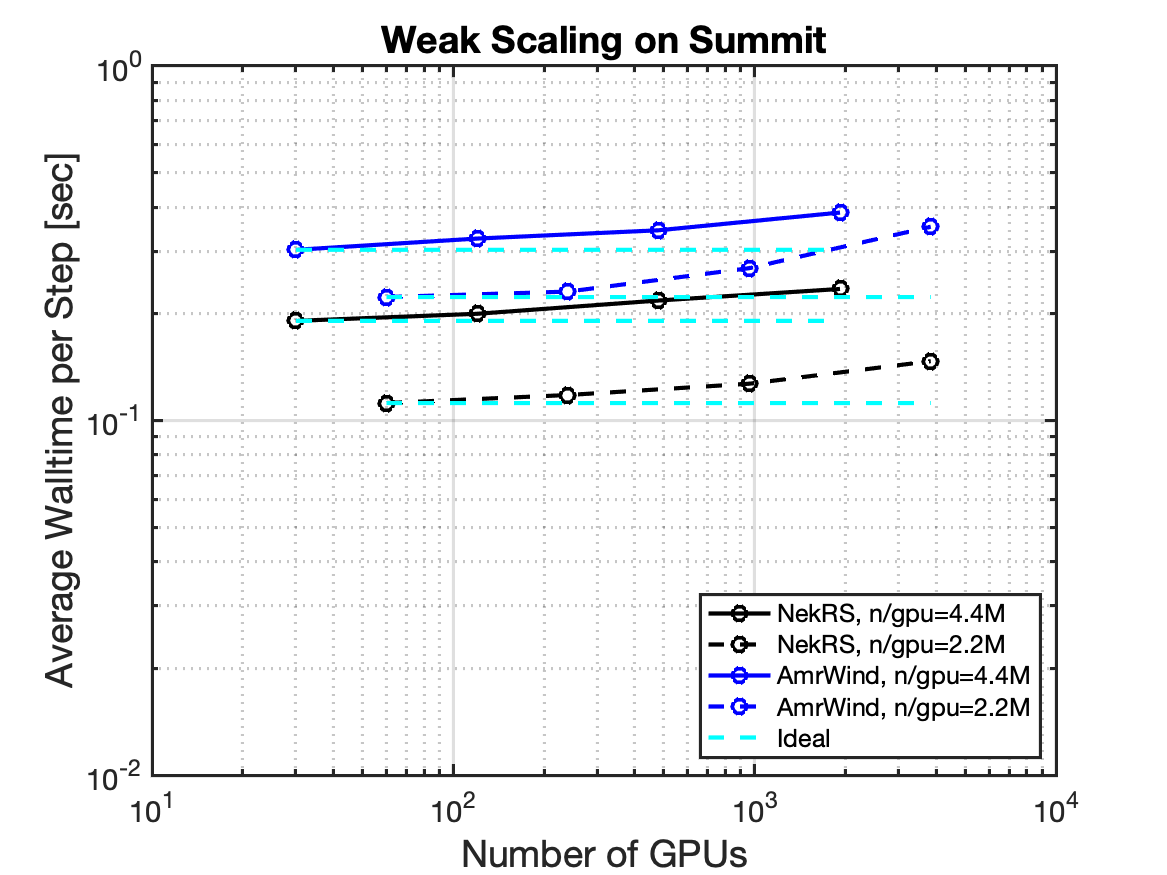}
    \hspace{-2em}
    \\
    \includegraphics[width=0.35\textwidth]{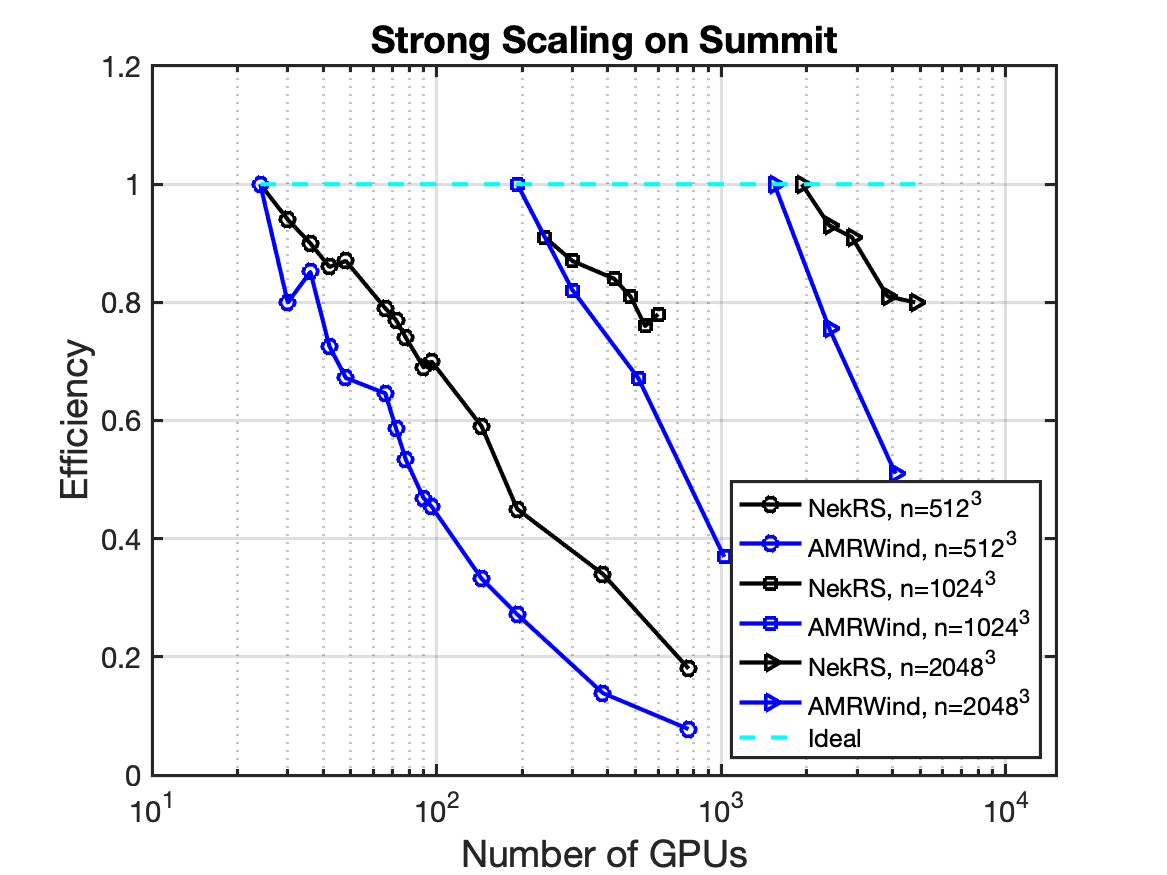}
    \hspace{-2em}
    \includegraphics[width=0.35\textwidth]{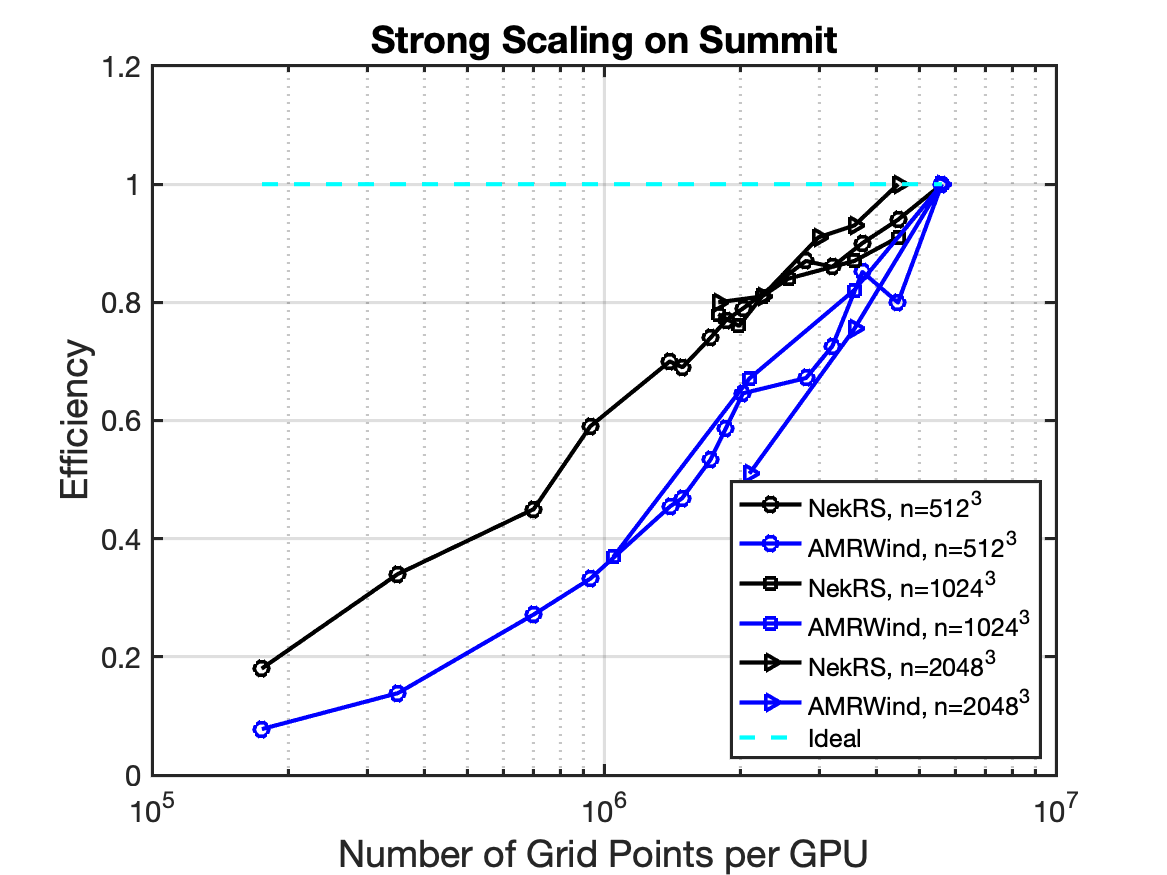}
    \hspace{-2em}
    \includegraphics[width=0.35\textwidth]{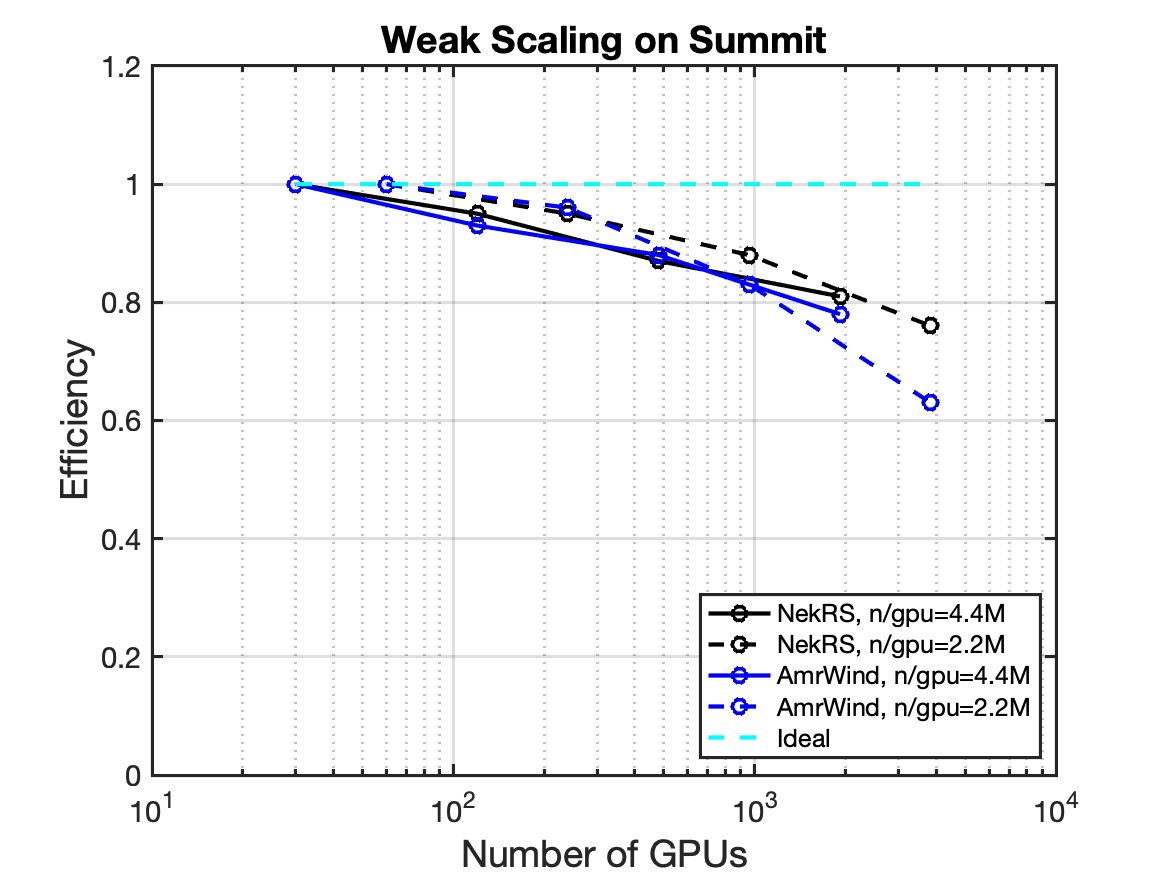}
    \hspace{-2em}
    \\
  \caption{\label{scale}NekRS vs. AMR-Wind strong and weak scaling on Summit GPUs.}
  \end{center}
\end{figure*}

\begin{figure*}[t]
  \begin{center}
     \includegraphics[width=0.35\textwidth]{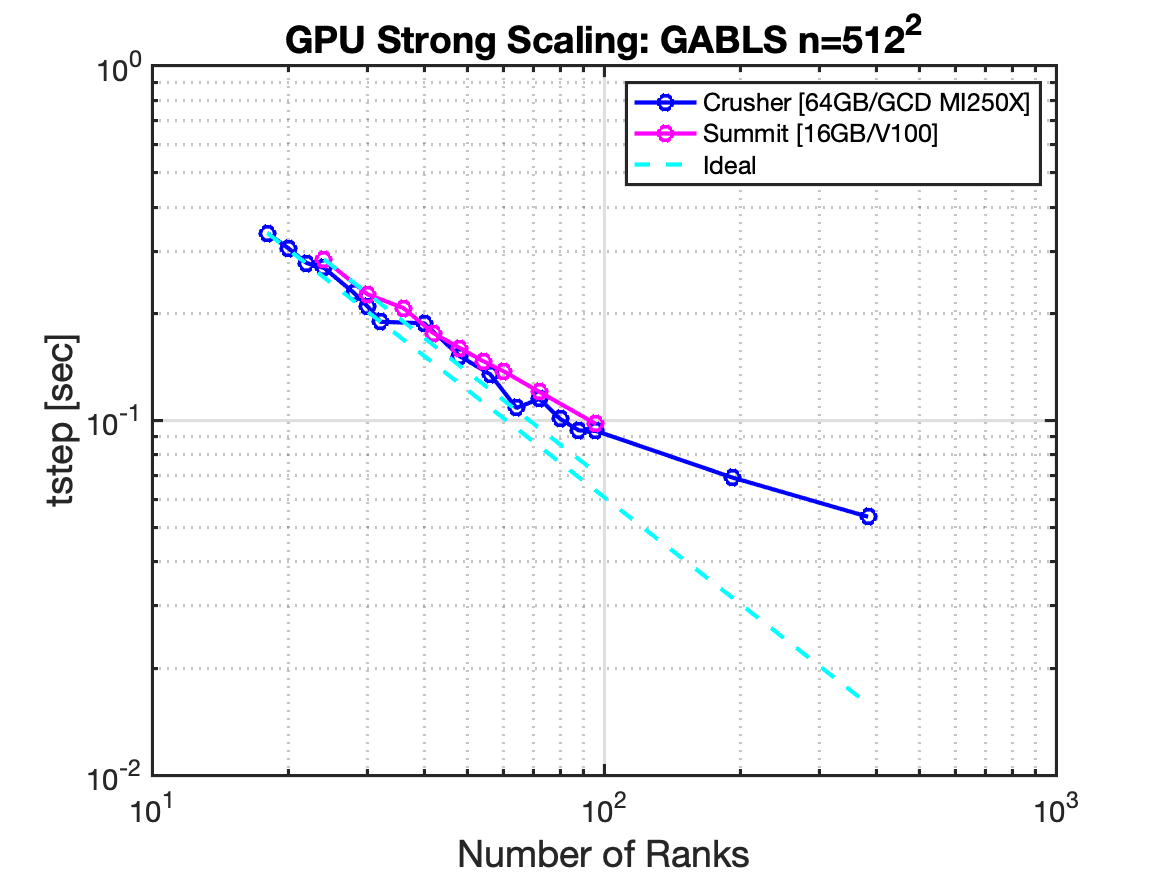}
     \hspace{-2em}
     \includegraphics[width=0.35\textwidth]{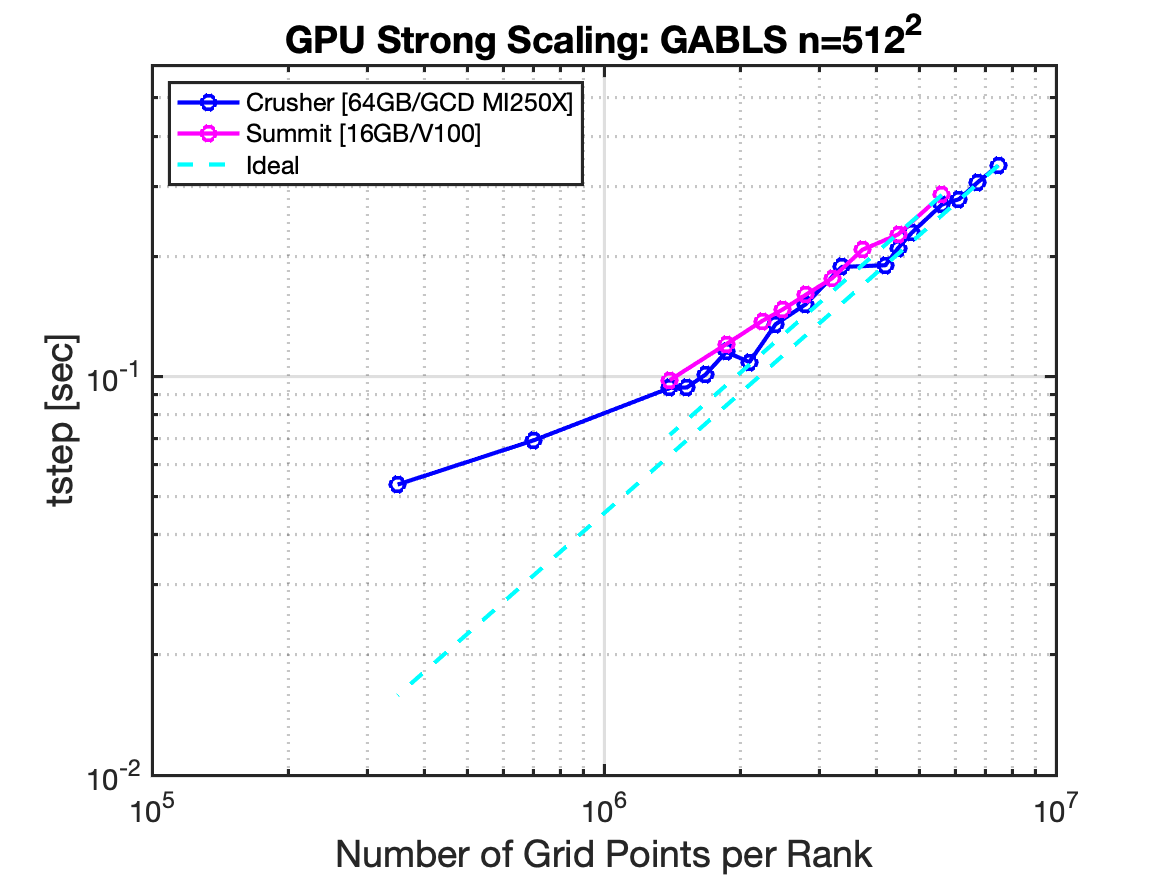}
     \\
     \includegraphics[width=0.35\textwidth]{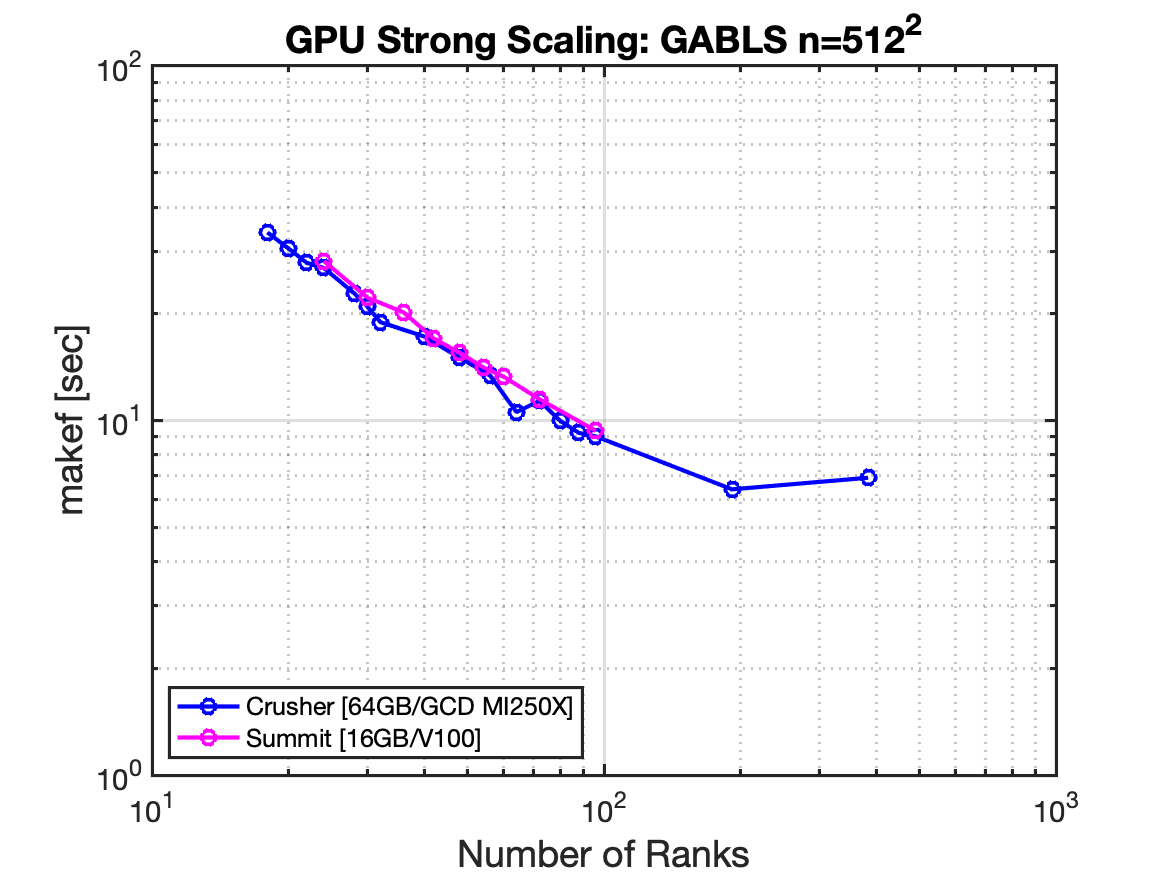}
     \hspace{-2em}
     \includegraphics[width=0.35\textwidth]{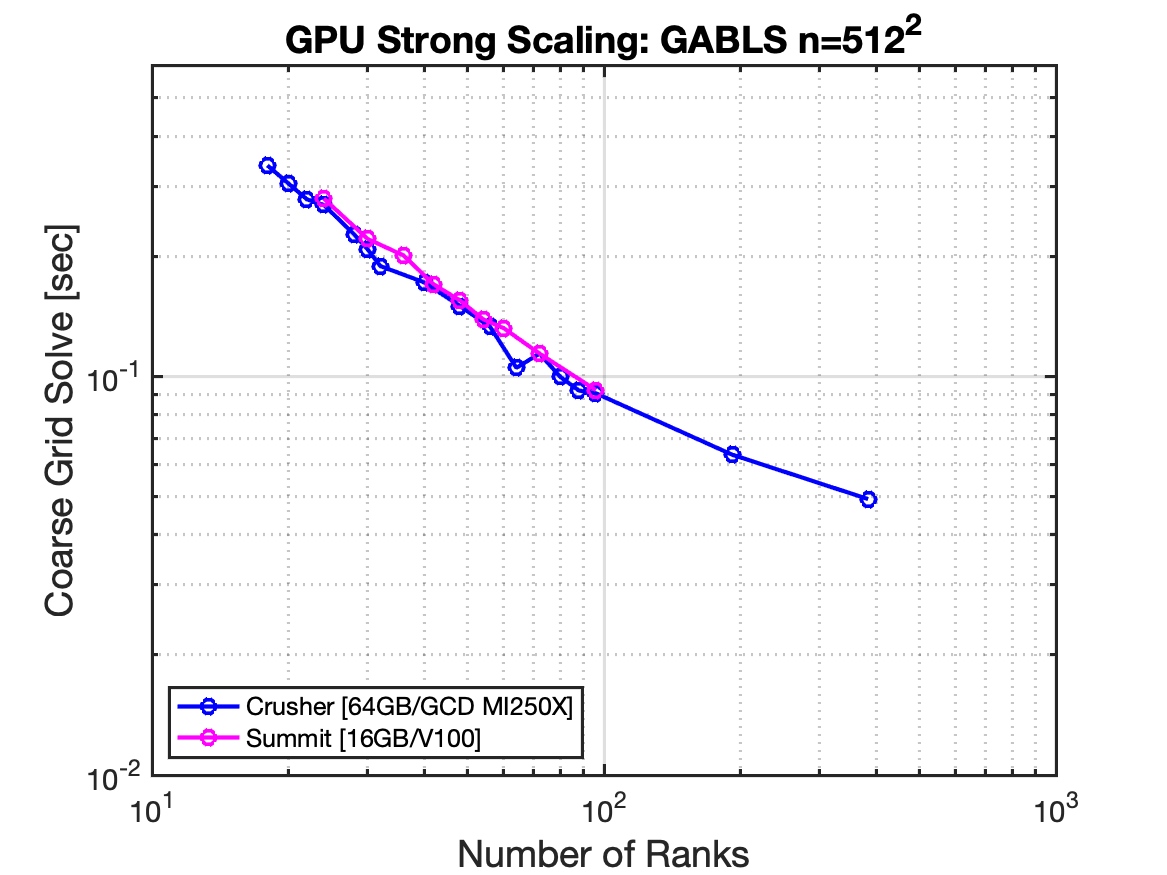}
   \caption{\label{perf10}NekRS GPU strong-scaling comparison on Crusher and Summit.}
  \end{center}
\end{figure*}

We next consider GPU-only performance on Summit using a single V100 per MPI
rank.  Figure~\ref{scale}, top, shows performance in terms of $t_{step}$ for
strong scaling as a function of the number of of GPUs, $P$, in the left column
and as a function of number of points per rank, $n/P$, in the center column.
Weak-scaling performance is presented in the right column.  The wall-time
figure also shows the ideal speed-up curves scaling as $P^{-1}$.    The lower
plots show parallel efficiency, \begin{eqnarray} P_{\mathrm{eff}} & := & \frac{t_{0}
P_{0}}{t_{step} P}, \end{eqnarray} where $P_0$ is the smallest value of $P$
that will hold the given problem and $t_0$ is the $t_{step}$ value
corresponding to $P_0$.

We see that at the lower resolution of $n=512^3$, the performance of the two
codes is within a factor of 2 of each other out to $P=78$.  From the
efficiency figures we can see that both curves have dropped below 80\%
efficiency by that point, so a more realistic point of comparison would be at
$P=66$ given that users would typically not run this relatively small case on
$P > 66$.  We note that $P=66$ corresponds to $n/P=2$M, which is a typical
strong-scaling limit for NekRS on current-generation GPU platforms.

 \begin{table*} [t]
  \footnotesize
  \begin{center}
  \begin{tabular}{|c|c|c||c|c|c|c|c|c||c|c|c|c|c|c|}
  \hline
   \multicolumn{15}{|c|}{{\bf  Strong-Scaling on Summit GPU}, $n=512^3$, $\Delta x=0.78$ m, $\Delta t=6.25$e-2s,  $\Omega=$ [400 m $\times$ 400 m $\times$ 400 m]} \\
   \hline
   \multicolumn{3}{|c||}{} &
   \multicolumn{6}{|c||}{NekRS} &
   \multicolumn{6}{|c|}{AMR-Wind} \\
   \hline
         node & gpu  & $n$/gpu & $v_i$  & $p_i$ &  $T_i$ & $t_{step}$  & $P_{\rm eff}$ & $r_t$ & $v_i$  & $p_i$ &  $T_i$ & $t_{step}$  & $P_{\rm eff}$ & $r_t$ \\
   \hline
   4   &  24   &   5.5924e+06 &   2  &  1.81 &   1 &   2.44e-01 &    100 &  3.90 & 2  &  2  &   2 &  3.19e-01 & 100 &    5.10\\
   8   &  48   &   2.7962e+06 &   2  &  1.82 &   1 &   1.39e-01 &    87  &  2.22 & 2  &  2  &   2 &  2.37e-01 &  67 &    3.80\\
   11  &  66   &   2.0336e+06 &   2  &  1.85 &   1 &   1.11e-01 &    79  &  1.78 & 2  &  2  &   2 &  1.79e-01 &  64 &    2.87\\
   16  &  96   &   1.3981e+06 &   2  &  1.90 &   1 &   8.66e-02 &    70  &  1.38 & 2  &  2  &   2 &  1.75e-01 &  45 &    2.80\\
   24  & 144   &   9.3207e+05 &   2  &  2.00 &   1 &   6.87e-02 &    59  &  1.09 & 2  &  2  &   2 &  1.60e-01 &  33 &    2.56\\
   32  & 192   &   6.9905e+05 &   2  &  2.00 &   1 &   6.77e-02 &    45  &  1.08 & 2  &  2  &   2 &  1.46e-01 &  27 &    2.34\\
   64  & 384   &   3.4953e+05 &   2  &  2.00 &   1 &   4.40e-02 &    34  &  0.70 & 2  &  2  &   2 &  1.43e-01 &  13 &    2.30\\
   128 & 768   &   1.7476e+05 &   2  &  2.00 &   1 &   4.02e-02 &    18  &  0.64 & 2  &  2  &   2 &  1.28e-01 &  7.7&    2.05\\
   256 & 1536  &   8.7381e+04 &   2  &  2.00 &   1 &   3.60e-02 &    10  &  0.57 & 2  &  2  &   2 &  1.41e-01 &  3.5&    2.26\\
  \hline         
  \hline         
  \hline
   \multicolumn{15}{|c|}{{\bf  Strong-Scaling on Summit GPU}, $n=1024^3$, $\Delta x=0.39$ m, $\Delta t=3.125$e-2s,  $\Omega=$ [400 m $\times$ 400 m $\times$ 400 m]} \\
   \hline
   \multicolumn{3}{|c||}{} &
   \multicolumn{6}{|c||}{NekRS} &
   \multicolumn{6}{|c|}{AMR-Wind} \\
   \hline
         node & gpu & $n$/gpu  & $v_i$  & $p_i$ &  $T_i$ & $t_{step}$  & $P_{\rm eff}$ & $r_t$ & $v_i$  & $p_i$ &  $T_i$ & $t_{step}$  & $P_{\rm eff}$ & $r_t$ \\
   \hline
   32  &  192    &  5.5924e+06   & 1  &  1.4  & 1  & 2.34e-01  &  100  & 7.50 &  2    &   2    &   2   &    0.369   &   100  &  11.82\\
   40  &  240    &  4.4739e+06   & 1  &  1.3  & 1  & 2.04e-01  &  91   & 6.54 &  2    &   2    &   2   &    0.402   &   73   &  12.86\\
   50  &  300    &  3.5791e+06   & 1  &  1.4  & 1  & 1.72e-01  &  87   & 5.50 &  2    &   2    &   2   &    0.288   &   82   &   9.21\\
   70  &  420    &  2.5565e+06   & 1  &  1.4  & 1  & 1.27e-01  &  84   & 4.06 &  2    &   2    &   2   &    0.303   &   56   &   9.72\\
   80  &  480    &  2.2370e+06   & 1  &  1.3  & 1  & 1.15e-01  &  81   & 3.68 &  2    &   2    &   2   &    0.301   &   49   &   9.65\\
   86  &  512    &  2.0972e+06   &    &   -   &    &   -       &   -   &      &  2    &   2    &   2   &    0.206   &   67   &   6.60\\
   90  &  540    &  1.9884e+06   & 1  &  1.3  & 1  & 1.08e-01  &  76   & 3.47 &  2    &   2    &   2   &    0.217   &   60   &   6.95\\
   100 &  600    &  1.7896e+06   & 1  &  1.3  & 1  & 9.57e-02  &  78   & 3.06 &  2    &   2    &   2   &    0.219   &   54   &   7.01\\
  \hline
  \hline
   \multicolumn{15}{|c|}{{\bf  Strong-Scaling on Summit GPU}, $n=2048^3$, $\Delta x=0.39$ m, $\Delta t=1.5625$e-2s,  $\Omega=$ [400 m $\times$ 400 m $\times$ 400 m]} \\
   \hline
   \multicolumn{3}{|c||}{} &
   \multicolumn{6}{|c||}{NekRS} &
   \multicolumn{6}{|c|}{AMR-Wind} \\
   \hline
         node & gpu  & $n$/gpu & $v_i$  & $p_i$ &  $T_i$ & $t_{step}$  & $P_{\rm eff}$ & $r_t$ & $v_i$  & $p_i$ &  $T_i$ & $t_{step}$  & $P_{\rm eff}$ & $r_t$ \\
  \hline
   256  & 1536    &  5.5924e+06 &   &  -    &    &      -    &      - &      &    2   &    2   &    2   &    0.437  &    100  &  34.99\\  
   320  & 1920    & 4.4739e+06  & 1 &  1.16  & 1 &  2.10e-01 &  100   & 16.8 &    2   &    2   &    2   &    0.485  &    72   &  38.80\\
   400  & 2400    & 3.5791e+06  & 1 &  1.20  & 1 &  1.80e-01 &  93    & 14.4 &    2   &    2   &    2   &    0.370  &    76   &  29.62\\
   480  & 2880    & 2.9826e+06  & 1 &  1.22  & 1 &  1.54e-01 &  91    & 12.3 &    2   &    2   &    2   &    0.402  &    58   &  32.16\\
   640  & 3840    & 2.2370e+06  & 1 &  1.25  & 1 &  1.28e-01 &  81    & 10.3 &    2   &    2   &    2   &    0.440  &    40   &  35.26\\
   683  & 4096    & 2.0972e+06  &   &    -   &   &    -      &   -    &      &    2   &    2   &    2   &    0.321  &    51   &  25.69\\
   800  & 4800    & 1.7896e+06  & 1 &  1.18  & 1 &  1.05e-01 &  80    & 8.4  &    2   &    2   &    2   &    0.390  &    36   &  31.23\\
  \hline
  \end{tabular}
 \end{center}
 \caption{\label{nek-abl-strong1} NekRS GPU vs. AMR-Wind GPU strong-scaling performance study.}
 \end{table*}

 \begin{table*} [t]
  \footnotesize
  \begin{center}
  \begin{tabular}{|c|c|c|c|c||c|c|c|c|c|c||c|c|c|c|c|c|}
  \hline
   \multicolumn{17}{|c|}{{\bf  Weak-Scaling on Summit GPU}, $\Delta x=0.78$ m, $\Delta t=6.25$e-2s} \\
   \hline
   \multicolumn{5}{|c||}{} &
   \multicolumn{6}{|c||}{NekRS} &
   \multicolumn{6}{|c|}{AMR-Wind} \\
   \hline
         node & gpu & $n$ & $\Omega$  & $n$/gpu & $v_i$  & $p_i$ &  $T_i$ & $t_{step}$  & $P_{\rm eff}$ & $r_t$ & $v_i$  & $p_i$ &  $T_i$ & $t_{step}$  & $P_{\rm eff}$ & $r_t$ \\
   \hline
   5 &  30  &  512$^2\times$ 512 & 400 m$^2\times$400 m &  4.4739e+06  & 1   & 1.5 & 1 & 0.191 & 100 & 3.8 & 2  & 2 & 2 & 0.303 & 100 & 4.8\\ 
  20 & 120  & 1024$^2\times$ 512 & 800 m$^2\times$400 m &  4.4739e+06  & 1   & 1.7 & 1 & 0.200 &  95 & 4.0 & 2  & 2 & 2 & 0.326 & 93  & 5.2\\ 
  80 & 480  & 2048$^2\times$ 512 &1600 m$^2\times$400 m &  4.4739e+06  & 1   & 1.9 & 1 & 0.218 &  87 & 3.8 & 2  & 2 & 2 & 0.344 & 88  & 5.5\\ 
 320 & 1920 & 4096$^2\times$ 512 &3200 m$^2\times$400 m &  4.4739e+06  & 1   & 2.4 & 1 & 0.235 &  81 & 4.7 & 2  & 2 & 2 & 0.386 & 78  & 6.2\\ 
 \hline
   \hline
   \multicolumn{5}{|c||}{} &
   \multicolumn{6}{|c||}{NekRS} &
   \multicolumn{6}{|c|}{AMR-Wind} \\
   \hline
         node & gpu  & $n$ & $\Omega$  & $n$/gpu  & $v_i$  & $p_i$ &  $T_i$ & $t_{step}$  & $P_{\rm eff}$ & $r_t$ & $v_i$  & $p_i$ &  $T_i$ & $t_{step}$  & $P_{\rm eff}$ & $r_t$ \\
   \hline
  10 &  60  &  512$^2\times$ 512 &  400 m$^2\times$400 m  & 2.2370e+06  & 1  & 1.5 &1 & 0.112 &  100 & 2.2 & 2  & 2 & 2 & 0.223 & 100 & 3.6\\ 
  40 & 240  & 1024$^2\times$ 512 &  800 m$^2\times$400 m  & 2.2370e+06  & 1  & 1.7 &1 & 0.118 &   95 & 2.4 & 2  & 2 & 2 & 0.231 & 96  & 3.7\\ 
 160 & 960  & 2048$^2\times$ 512 & 1600 m$^2\times$400 m  & 2.2370e+06  & 1  & 2.1 &1 & 0.127 &   88 & 2.5 & 2  & 2 & 2 & 0.269 & 83  & 4.3\\ 
 640 & 3840 & 4096$^2\times$ 512 & 3200 m$^2\times$400 m  & 2.2370e+06  & 1  & 2.5 &1 & 0.147 &   76 & 2.9 & 2  & 2 & 2 & 0.352 & 63  & 5.6\\ 
  \hline
  \end{tabular}
  \end{center}
 \caption{\label{nek-abl-weak1} NekRS GPU vs. AMR-Wind GPU weak-scaling performance study with fixed mesh density and resolution per GPU. 
  }
 \end{table*}


The center column in Fig.~\ref{scale} replots the strong-scaling information
with $n/P$ as the independent variable.  Here we see a collapse of each code's 
strong-scale data into a single curve, particularly for NekRS.
The efficiency plot, lower-center, clearly shows the $n/P=2$M mark as
the 80\% parallel efficiency point for NekRS.   The AMR-Wind wall-time
curves, upper center, are not as tightly grouped, particularly for the
large problem sizes on large processor counts.  It is tempting to speculate
that this increased cost is due to an increase in iteration count, but 
Tables \ref{nek-abl-strong1} and \ref{nek-abl-weak1} show that is not the
case, since each solver requires only two iterations per timestep for each of the
problems.  An important feature of AMR-Wind is that it generally performs
better if $P$ is a power of 2.  
At the critical point of $n/P=2$M, NekRS is only a factor of 1.6
faster than AMR-Wind for the $n=512^3$ case.   

Figure~\ref{scale}, right, shows weak-scaling results for $n/P=2.2$M and 4.4M. 
For the heavily loaded cases, AMR-Wind is within a factor of 1.6 of NekRS, but this
figure increases to roughly a factor of 2 for the 2.2M points-per-GPU case.
The weak-scale efficiency reaches 80\% at around $P=2000$ GPUs for all the cases
save the AMR-Wind case with $n/P=2.2$M, which crosses the 80\% mark at
$P \approx 1100$.

Tables \ref{nek-abl-strong1} and \ref{nek-abl-weak1} provide a detailed breakdown
of several of the key metrics for the code performance, including iteration counts
($v_i$, $p_i$, $T_i$, for the respective velocity, pressure, and temperature iterative
solvers), $t_{step}$, parallel efficiency ($P_{\rm eff}$), and the wall-time to
physical-time ratio ($r_t$). This last quantity is of particular interest since 
it must be smaller than unity for weather modeling applications. 
We also note that $P$ is denoted by gpu in the tables.  We see from Table
\ref{nek-abl-strong1} that, for a fixed value of $n/P$, $r_t$ effectively
doubles with each doubling of (linear) resolution.  The reason for this
increase is that the number of timesteps must also double whenever the
number of points in each direction is doubled (for fixed domain size).
Throughout the table, we see that roughly two iterations are required per
timestep for each of the linear solvers, indicating that the preconditioners
are robust with respect to mesh size, although NekRS does show some
increase in iteration count in the weak-scale results. 
   
We remark that AMR supports block-structured adaptive mesh refinement, which
means that static grids do not leverage one of its main features.  It is
nonetheless highly performant on this problem.  Moreover, AMR-Wind has a
significant performance boost when the number of ranks is a power of 2, as seen
in Table \ref{nek-abl-strong1} for the $n=1024^3$ case for $P=512$ and in the
$n=2048^3$ case for $P=4096$.  In the former case, the parallel efficiency
jumps from 49\% to 67\% as $P$ changes from 480 to 512.  In the latter, it jumps
from 40\% for $P=3840$ to 51\% for $P=4096$.   These performance gains derive
from the block decompositions used in AMR-Wind, which favor block sizes 
(and thus, processor counts) that are powers of 2.  

We close with a scaling comparison of Summit and Crusher performance
for NekRS in Fig. \ref{perf10}.  The upper figures show standard strong scaling
as a function of the number of ranks on the left (one GPU or GCD per rank) and
as a function of $n/P$ on the right.  The lower plots show the timing for the
{\tt makef} kernel (left), which evaluates the nonlinear advection term and
does not require communication, and for the coarse-grid solve (right), which is
communication dominated.  The coarse-grid problem, which has roughly $E$
degrees of freedom (with $E=262144$ in this case), is solved by using algebraic
multigrid (hypre) on the host CPUs.  The performance for these two platforms is
remarkably similar.


 \section{Conclusion}

We presented profiling and timing results for two CFD codes, NekRS and
AMR-Wind, applied to the GABLS atmospheric boundary layer test problem, which
is of direct relevance to wind farm modeling and weather forecasting.  Strong
and weak scaling were demonstrated on up to $P=4800$ NVIDIA V100 GPUs on OLCF's
Summit.  For NekRS, wall-clock times of 0.11 s were observed for $n/P=2$M,
which is the 80\% efficiency point across a range of problem sizes.  AMR-Wind
was generally within a factor of 1.4--2.0 of the performance of NekRS over the
range of interest.    For both codes, the substep that inhibits strong scaling
is the intrinsically communication-intensive pressure Poisson solve.   For
NekRS it was shown that a single GCD of the MI250X on Crusher is delivering
performance that is comparable to a single V100 on Summit.
Finally, we demonstrated that careful subgrid-scale modeling is critical to
realizing comparable results.  A future paper will investigate the modeling
questions more deeply.

\section*{Acknowledgments}

This material is based upon work supported by the U.S. Department of Energy, 
Office of Science, under contract DE-AC02-06CH11357 and by the Exascale Computing 
Project (17-SC-20-SC). The research used resources at the Oak Ridge Leadership 
Computing Facility at Oak Ridge National Labo- ratory, which is supported by 
the Office of Science of the U.S. Department of Energy under Contract DE-AC05-00OR22725.



\bibliographystyle{./SageH}
 \bibliography{bibs/emmd,bibs/ananias,bibs/abl,bibs/references,bibs/amr}

\end{document}